\documentclass[aps,twocolumn,pre,amsmath,amssymb,floatfix,eqsecnum]{revtex4-1}
\usepackage{color}

\usepackage{graphicx,epsfig,bm}

\newcommand{\tb}{\mathring{\tau}}

\newcommand{\ub}{\mathring{u}}

\newcommand{\rmd}{\text{d}}
\newcommand{\rme}{\text{e}}
\newcommand{\rmi}{\text{i}}
\newcommand{\psih}{\hat{\psi}}
\newcommand{\lamth}{\lambda_{\text{th}}}
\newcommand{\ef}{\mathfrak{h}}
\newcommand{\ev}{\varepsilon}
\newcommand{\Li}{\text{Li}}

\newcommand{\pint}{\int_{\bm{p}}^{(d-1)}} \newcommand{\psum}{\frac{1}{A}\sum_{\bm{p}}}

\newcommand{\Tr}{\mathop{\text{Tr}}}

\newcommand{\tc}{T_{\text{c}}}
\newcommand{\arsinh}{\mathop{\text{arsinh}}}

\newcommand{\Htheta}{\theta}

\renewcommand{\Re}{\text{Re}} \renewcommand{\Im}{\text{Im}}

\newcommand{\sk}{\mathsf{k}}

\newcommand{\scc}{\mathsf{c}}

\definecolor{dblue}{rgb}{0.1,0.1,0.44}
\definecolor{dgreen}{rgb}{0.2 ,0.54, 0.2}

\newcommand{\mat}[1]{\mathbf{#1}}

\newcommand{\Or}{\text{O}}

\newcommand{\aseq}{\simeq}
\newcommand{\asprop}{\sim}
\newcommand{\etwa}{\approx}

\begin{document}

\title{Fluctuation-induced forces in confined ideal and imperfect Bose gases}

\author{H.~W. Diehl and Sergei B.\ Rutkevich}
\affiliation{Fakult\"at f\"ur Physik, Universit\"at Duisburg-Essen, D-47058 Duisburg, Germany}

\date{\today}

\begin{abstract}
Fluctuation-induced (``Casimir'') forces caused by thermal and quantum fluctuations are investigated for ideal and imperfect Bose gases confined to $d$-dimensional films of size $\infty^{d-1}\times D$ under periodic (P), antiperiodic (A), Dirichlet-Dirichlet (DD), Neumann-Neumann (NN), and Robin (R) boundary conditions (BCs). The full scaling functions $\Upsilon^{\text{BC}}_d({x_\lambda=D/\lamth},{x_\xi=D/\xi})$ of the residual reduced grand potential per area, $\varphi_{\text{res},d}^{\text{BC}}(T,\mu,D)=D^{-(d-1)}\Upsilon_d^{\text{BC}}(x_\lambda,x_\xi)$, are determined for the ideal gas case with these BCs, where $\lamth$ and $\xi$ are the thermal de-Broglie wavelength and the bulk correlation length, respectively. The associated limiting scaling functions $\Theta^{\text{BC}}_d(x_\xi)\equiv \Upsilon_d^{\text{BC}}(\infty,x_\xi)$  describing the critical behavior at the bulk condensation transition  are shown to agree  with those previously determined from a massive free $O(2)$ theory for $\text{BC}=\text{P},\text{A},\text{DD},\text{DN},\text{NN}$. For $d=3$, they are expressed in closed analytical form in terms of polylogarithms.  The analogous scaling functions $\Upsilon_d^{\text{BC}}(x_\lambda,x_\xi,c_1D,c_2D)$ and $\Theta^{\text{R}}_d(x_\xi,c_1D,c_2D)$ under the RBCs $(\partial_z-c_1)\phi|_{z=0}=(\partial_z+c_2)\phi|_{z=D}=0$ with $c_1\ge 0$ and $c_2\ge 0$ are also determined. The corresponding scaling functions $\Upsilon_{\infty,d}^{\text{P}}(x_\lambda,x_\xi)$ and $\Theta_{\infty,d}^{\text{P}}(x_\xi)$ for the imperfect Bose gas are shown to agree with those of the interacting Bose gas with $n$ internal degrees of freedom in the limit $n\to\infty$. Hence, for ${d=3}$, $\Theta_{\infty,d}^{\text{P}}(x_\xi)$ is known exactly in closed analytic form. To account for the breakdown of translation invariance in the direction perpendicular to the boundary planes implied by free BCs such as DDBCs, a modified imperfect Bose~gas model is introduced that corresponds to the  limit $n\to\infty$ of this interacting Bose gas. Numerically and analytically exact results for the scaling function $\Theta_{\infty,3}^{\mathbb{DD}}(x_\xi)$ therefore follow from those of the  $O(2n)$ $\phi^4$ model  for $n\to\infty$. \end{abstract}

\maketitle

\section{Introduction \label{sec:intro}}

When a macroscopic system consisting of a medium in which long-wavelength low-energy excitations can occur is confined along a given direction, fluctuation-induced effective forces can emerge. These fluctuations can be of  quantum mechanical  or classical (i.e., thermal) nature. An example of the first kind of fluctuation-induced forces are the Casimir forces \cite{Cas48} between two grounded parallel metallic plates caused by  the modification of the vacuum fluctuations of the electromagnetic field due to the presence of the plates. Familiar examples of fluctuation-induced forces of the second kind are the critical Casimir forces  that appear near continuous phase transitions with a bulk critical temperature $T_c>0$ \cite{Kre94,BDT00,Gam09}. 

In confined quantum systems, generally both quantum and thermal fluctuations occur. At a conventional critical point with $T_c>0$, quantum fluctuations are expected to be irrelevant, i.e., they give only corrections to the leading asymptotic behavior on large length scales \cite{Her76,Sac11}. However, at sufficiently low temperature or near quantum critical points, quantum fluctuations are crucial and must not be neglected. An important prototype class of systems exhibiting both quantum and thermal fluctuations are Bose gases. In this paper we are concerned with fluctuation-induced forces of ideal and interacting Bose gases confined to a hypercuboid of size $A\times D$ of  cross-sectional hyperarea $A=L^{d-1}\to\infty$ and finite width $D$ in dimensions $2<d<4$.

Consider first the ideal Bose~gas case. For $d=3$ this has been investigated in some detail in \cite{MZ06} for the cases of periodic (P), Dirichlet-Dirichlet (DD), and Neumann-Neumann (NN) boundary conditions (BCs) along the finite direction and chemical potentials $\mu<\mu_{\text{c,id}}$, where $\mu_{\text{c,id}}=0$ is the bulk ($D=\infty$)  critical value of $\mu$ at the Bose-Einstein transition \cite{fn11,Bis07}. Let 
\begin{equation}
\varphi_d^{\text{BC}}(T,\mu,D)=-\lim_{L\to\infty}L^{1-d}\ln \Xi^{\text{BC}}_d(T,\mu,D,L)
\end{equation}
be the reduced grand potential per cross-sectional area $A$, where $\Xi^{\text{BC}}_d$ denotes the grand partition function and the superscript $\text{BC}$ indicates the type of boundary conditions chosen along the finite direction, e.g., $\text{BC}=\text{P}, \text{DD}, \text{NN},\text{DN}$. We will also consider antiperiodic ($\text{BC}=\text{A}$) and Robin BC ($\text{BC}=\text{R}$) \cite{Die86a,SD08,DS11}. The latter will be specified below.

Writing
\begin{eqnarray}
\varphi_d^{\text{BC}}(T,\mu,D)&=&D\varphi_{\text{b},d}(T,\mu)+\varphi_{\text{s},d}^{\text{BC}}(T,\mu)\nonumber\\ &&\strut+\varphi_{\text{res},d}^{\text{BC}}(T,\mu,D),
\end{eqnarray}
we decompose this  reduced grand potential  into a contribution involving the reduced bulk potential $\varphi_{\text{b},d}$, a $D$-independent surface term $\varphi_{\text{s},d}^{\text{BC}}$, and a $D$-dependent remainder $\varphi_{\text{res},d}^{\text{BC}}$,  which we call residual reduced potential \cite{fn1,JNS16}. For the ideal Bose gas, one has the well-known result (see, e.g., \cite{GB68})
\begin{equation}\label{eq:varphib}
\varphi_{\text{b},d}(T,\mu<0)=-\lamth^{-d}\,\Li_{d/2+1}\big(\rme^{\beta\mu}\big).
\end{equation}

Aside from the width $D$, there are two lengths in the problem \cite{fn2,WRFS86}. One is  the thermal de-Broglie wavelength
\begin{equation}
\lamth=\hbar \sqrt{2\pi \beta /m},
\end{equation}
where $m$ is the mass of the Bose particles and $\beta=(k_BT)^{-1}$. The other is the bulk correlation length $\xi$, which in the ideal Bose~gas case is given by
\begin{equation}\label{eq:xiid}
 \xi_{\text{id}}=\frac{\hbar}{\sqrt{2m(-\mu)}}.
\end{equation}
The latter is finite or infinite depending on whether $\mu<0$  or  $\mu=0$. On dimensional grounds,  the residual potential  can therefore be written in the form
\begin{equation}\label{eq:fresUp}
\varphi_{\text{res},d}^{\text{BC}}(T,\mu,D)=D^{-(d-1)}\,\Upsilon_d^{\text{BC}}(D/\lamth,D/\xi).
\end{equation}

Let us also introduce the reduced Casimir pressure
\begin{equation}
\beta\mathcal{F}^{\text{BC}}_C=-\frac{\partial}{\partial D}\varphi_{\text{res},d}^{\text{BC}}(T,\mu,D).
\end{equation}
This can be written in the analogous scaling form
\begin{equation}\label{eq:betaFC}
\beta\mathcal{F}^{\text{BC}}_C(T,\mu,D)=D^{-d}\,\mathcal{Y}_d^{\text{BC}}(D/\lamth,D/\xi),
\end{equation}
whose scaling function can be expressed in terms of $\mathcal{Y}_d^{\text{BC}}$ and its derivatives as
\begin{equation}\label{eq:Ydrel}
\mathcal{Y}_d^{\text{BC}}(x_\lambda,x_\xi)=\bigg(d-1-x_\lambda\frac{\partial}{\partial x_\lambda}-x_\xi\frac{\partial}{\partial x_\xi}\bigg)\Upsilon_d^{\text{BC}}(x_\lambda,x_\xi).
\end{equation}

In \cite{MZ06}, the ideal~Bose gas functions $\Upsilon_{{d=3}}^{\text{BC}}$ were computed  at ${d=3}$  for the bulk disordered phase $\mu<0$ in the form of double series. The asymptotic scaling forms of the functions $\varphi_{\text{res},d}^{\text{BC}}$ for $\xi\to\infty$ are expected to be independent of quantum effects. The dependence on the thermal de Broglie wavelength  $\lamth$ should drop out in the limit $D/\lamth\to\infty$,  so that Eqs.~\eqref{eq:fresUp} and \eqref{eq:betaFC} must  asymptotically reduce to
 \begin{equation}\label{eq:UpsThetarel}
\varphi_{\text{res},d}^{\text{BC}}(T,\mu,D)\mathop{\aseq}\limits_{\lamth\ll D}D^{-(d-1)}\Theta_{d}^{\text{BC}}(D/\xi)
\end{equation}
and
\begin{equation}
\beta\mathcal{F}^{\text{BC}}_C(T,\mu,D)\mathop{\aseq}\limits_{\lamth\ll D}D^{-d}\vartheta_{d}^{\text{BC}}(D/\xi)
\end{equation}
with
\begin{equation}\label{eq:ThetaBCdef}
\Theta_{d}^{\text{BC}}(x_\xi)=\Upsilon_d^{\text{BC}}(\infty,x_\xi)
\end{equation}
and
\begin{equation}
\label{eq:varhetaBCdef}
\vartheta_{d}^{\text{BC}}(x_\xi)=\mathcal{Y}_d^{\text{BC}}(\infty,x_\xi),
\end{equation}
where 
\begin{equation}
\vartheta_{d}^{\text{BC}}(x_\xi)=(d-1)\,\Theta_{d}^{\text{BC}}(x_\xi)-x_\xi\,\frac{\rmd}{\rmd x_\xi}\Theta_{d}^{\text{BC}}(x_\xi)
\end{equation}
as a consequence of Eq.~\eqref{eq:Ydrel}.

As pointed out in  \cite{GD06a}, the results of \cite{MZ06}, when appropriately normalized,  confirm this expectation and yield the functions $\Theta_{d}^{\text{BC}}(x_\xi)$ for $d=3$ and PBCs, DDBCs, and NNBCs in the form of series. The authors of  \cite{GD06a} furthermore showed that these functions agree with those  previously determined  for a free Gaussian theory with an $n$-component real-valued order parameter $\bm{\phi}$ \cite{KD92a} for the choice $n=2$.

Following, we generalize these ideal Bose gas results in four ways:

(i) In addition to PBCs, DDBCs, and NNBCs, we also consider DNBCs, ABCs, and RBCs \cite{Die86a,SD08,DS11,EGJK08}.  For the latter  the eigenfunctions $\ef(z)$ of the  operator $-\partial_z^2$ are required to satisfy along the finite $z$~direction the BCs 
\begin{equation}\label{eq:BCR}
(\partial_z-c_1)\ef|_{z=0}=0=(\partial_z +c_2)\ef|_{z=D}
\end{equation}
at the boundary planes ${z=0}$ and $D$, respectively,  where $c_1$ and $c_2$ are general non-negative parameters.  In this case, the scaling functions $\Upsilon_d^{\text{R}}$ and $\Theta_d^{\text{R}}$ also depend on the two additional scaling variables $c_1D$ and $c_2D$. 

(ii) We show the equality of  $\Upsilon_d^{\text{BC}}(\infty,x_\xi)$ with the classical $n=2$ free field scaling functions $\Theta_{d}^{\text{BC}}(x_\xi)$ for all $d>2$ and $\text{BC}=\text{P},\text{A}, \text{DD},\text{DN}, \text{NN}$.

(iii) Summing the series of the functions $\Theta_{3}^{\text{BC}}(x)$ for $\text{BC}=\text{P},\text{A},\text{DD},\text{DN}, \text{NN}$, we derive closed  analytical expressions for them in terms of polylogarithms.

(iv) Finally,  we determine $\Upsilon_d^{\text{R}}(D/\lamth,D/\xi,c_1D,c_2D)$  for all $d>2$, generalizing previous $\xi=\infty$  results to $\xi<\infty$ \cite{SD08,DS11,DrFSchmidt}.

Note that unless we state the contrary, we will  restrict ourselves in our analysis of the ideal Bose gas to the bulk disordered phase.  Since for $d\in (2,3]$ a phase with long-range order is not possible for $T>0$ and $D<\infty$, this is not a severe restriction. 

In extending our analysis to the interacting Bose~gas case, we consider a model which  is called \emph{imperfect Bose gas} according to common, though debatable, terminology \cite{Dav72,NJN13}. This is a model for a gas of interacting bosons in a region $\mathfrak{V}$ whose interaction energy is approximated by  $a\hat{N}^2/(2V)$, where $\hat{N}$ is the number operator.  In a recent paper \cite{NJN13} a $d$-dimensional model of such an imperfect Bose gas confined to a hypercuboid of size $L^{d-1}\times D$ and subject to periodic boundary conditions was investigated. Considering the appropriate thermodynamic limits $L,D\to\infty$ and $L\to \infty$ at fixed $D<\infty$, the authors derived expressions for the bulk and residual grand potentials. They found that the critical exponents that characterize the critical behavior at the transition at fixed chemical potential for all $d>2$ dimensions agree with those of the spherical model. This prompted them to conclude that the universality class of the critical behavior at the condensation transition of the imperfect Bose gas is represented by the spherical model. On the other hand, they found that at ${d=3}$ the Casimir amplitude $\Delta_C$ associated with the residual free energy at the critical point  takes twice the value of its analog for the spherical model \cite{Dan96,Dan98}.

We clarify this issue by showing that the bulk, surface, and residual potentials of the imperfect Bose gas agree with the corresponding quantities for a system of interacting bosons with $n$ internal degrees of freedom $\alpha$  and a pair potential of the form $(\ub/n)\,\delta_{\alpha\beta}\,\delta(\bm{x}-\bm{x}')$ in the limit $n\to\infty$. This latter model, henceforth called \emph{$n$-component interacting Bose gas}, is defined in detail in the next section.  Using a coherent-state functional-integral formulation, we show that the leading asymptotic behavior near  its bulk condensation transition is described by the ${n=\infty}$ analogs $\Theta_{\infty,d}^{\text{P}}$ of the above scaling functions $\Theta_{d}^{\text{P}}$ of the classical $O(2n)$ real-valued $\phi^4$ theory. This means, in particular, that for ${d=3}$ the critical behavior of the imperfect Bose gas with PBCs is described by twice the exactly known scaling function of the mean spherical model \cite{Dan96,Dan98}. 

In view of the equivalence of the imperfect Bose gas with the interacting $n$-component Bose gas in the limit $n\to\infty$, it is natural to ask  whether generalizations of the former model  to other BCs such as free ones can be defined so that  this equivalence prevails. We introduce  such an imperfect Bose gas model with free BCs in Sec.~\ref{sec:ImBggen}. Its  scaling function $\Theta_{\infty,3}^{\text{DD}}(tD)$ may be obtained for all values of the scaling field $t\propto \mu_{\text{c}}-\mu\gtreqless 0$ from the numerical solution of the $O(\infty)$ $\phi^4$ model  determined in \cite{DGHHRS12,DGHHRS14,DBR14,DGHHRS15}. Furthermore, a variety of  exact analytical results may be inferred from those known for the $O(\infty)$ $\phi^4$ model subject to DDBCs \cite{BM77a,BM77c,DR14,RD15a,RD15b,DR17}. 

The remainder of this paper is divided into  four additional sections and three appendixes. Section~\ref{sec:models} serves to define the ideal, imperfect, and interacting Bose~gas models with which we are concerned, including their BCs, and to recall their coherent-space functional-integral representations. In Sec.~\ref{eq:scfcideal} our results for the scaling functions of the ideal Bose gas in the disordered bulk phase and their quantum corrections are presented. Section~\ref{sec:intBG} deals with the ${n\to\infty}$ limit of the  $n$-component Bose gas, the equivalent imperfect Bose gas, and their scaling functions for PBCs and DDBCs. The exact scaling functions $\Theta^{\text{P}}_{\infty,3}$ and $\vartheta^{\text{P}}_{\infty,3}$ are determined in closed analytic forms and shown to coincide with those of the $O(\infty)$ $\phi^4$ model up to a factor of $2$. For the case of DDBC, a number of exact properties of the scaling functions $\Theta^{\text{DD}}_{\infty,3}$ and $\vartheta^{\text{DD}}_{\infty,3}$ are deduced from known exact results for the latter classical model. Finally, Sec.~\ref{sec:sumconc} contains a brief summary of our results and concluding remarks.

\section{Ideal, imperfect, and interacting Bose gases\label{sec:models}}
We consider an  interacting  Bose gas described by the Hamiltonian
\begin{equation}\label{eq:Ham1}
\hat{H}=T[\psih^\dagger,\psih]+U[\psih^\dagger,\psih],
\end{equation}
where 
\begin{equation}\label{eq:T1}
T[\psih^\dagger,\psih]=\frac{\hbar^2}{2m}\int_{\mathfrak{V}}\rmd^dx\,[\nabla\psih^\dagger(\bm{x})]\cdot\nabla\psih(\bm{x})
\end{equation}
and 
\begin{eqnarray}\label{eq:U1}
U[\psih^\dagger,\psih]&=&\frac{1}{2}\int\limits_{\mathfrak{V}\times\mathfrak{V}}\rmd^dx\,\rmd^dx'\big[\psih^\dagger(\bm{x})\psih^\dagger(\bm{x}')\nonumber\\ &&\times u(\bm{x}-\bm{x}')\psih(\bm{x}')\psih(\bm{x})\big]
\end{eqnarray}
denote the operators of the kinetic and potential energy, respectively. The integration region $\mathfrak{V}$ is the $d$-dimensional hypercuboid $[0,L]^{d-1}\times[0,D]$. We write $\bm{x}=(\bm{y},z)$ and  choose PBCs along the first $d-1$ Cartesian directions $\bm{y}$. Along the remaining $z$~direction we  consider  PBCs, ABCS, DDBCs, NNBCs, DNBCs, and RBCs. From a physical point of view, DDBCs and RBCs are the most relevant BCs; they correspond to the cases of bosons confined along the $z$~direction by infinitely or finitely high potential barriers, respectively. PBs are of interest because they are a preferred choice in numerical analyses. The remaining BCs (ABCS, NNBCs, and DNBCs) are mainly of theoretical interest, though some of them have been considered in the literature \cite{MZ06}; whether and how they can be realized in experiments is not clear to us. 

The operators $\psih(\bm{x})$ and $\psih^\dagger(\bm{x})$ are Bose field annihilation and creation operators which for ${L=\infty}$ can be expressed as
\begin{equation}
\hat{\psi}^{\text{BC}}(\bm{y},z)=\sum_k\ef_k^{\text{BC}}(z)\int\frac{\rmd^{d-1}{p}}{(2\pi)^{(d-1)/2}}\,\rme^{\rmi \bm{p}\cdot\bm{y}}b_{\bm{p},k}
\end{equation}
in terms of Bose annihilation and creation operators satisfying the  commutation relations 
\begin{eqnarray}
[b_{\bm{p},k},b^\dagger_{\bm{p}',k'}]&=&\delta(\bm{p}-\bm{p}')\,\delta_{k,k'},\nonumber \\{}
[b_{\bm{p},k},b_{\bm{p}',k'}]&=&[b^\dagger_{\bm{p},k},b^\dagger_{\bm{p}',k'}]=0.
\end{eqnarray}
The $\ef_k^{\text{BC}}(z)$ are the orthonormalized  eigenfunctions with eigenvalues $(k^{\text{BC}})^2$ of the operator $-\partial_z^2$ for the specified BCs. Explicitly, one has \cite{KD92a,BDT00,SD08,DS11}
\begin{subequations}
\begin{equation}\label{eq:per}
\ef_k^{\text{P}}(z)=\frac{1}{\sqrt{D}}\rme^{\rmi kz},\;k=\frac{2\pi}{D}\,\nu,\;\nu\in\mathbb{Z},
\end{equation}
\begin{equation}\label{eq:ap}
\ef_k^{\text{A}}(z)=\frac{1}{\sqrt{D}}\rme^{\rmi kz},\;k=\frac{2\pi}{D}(\nu+\tfrac{1}{2})\;\nu\in\mathbb{Z},
\end{equation}
\begin{equation}\label{eq:DD}
\ef_k^{\text{DD}}(z)=\sqrt{\frac{2}{D}}\sin(kz),\;k=\frac{\pi}{D}\,\nu,\;\nu=1,2,\dots,\infty,
\end{equation}
\begin{equation}\label{eq:NN}
\ef_k^{\text{NN}}(z)=\begin{cases}D^{-1/2},&k=\nu=0,\\ \sqrt{2/D}\cos(kz),&k=\frac{\pi}{D}\,\nu,\;\;\nu=1,2,\dotsc,\infty,
\end{cases}
\end{equation}
\begin{equation}\label{eq:DN}
\ef_k^{\text{DN}}(z)=\sqrt{\frac{2}{D}}\sin(kz),\;k=\frac{\pi}{D}(\nu+\tfrac{1}{2}),\;\nu=0,1,2,\dots,\infty,
\end{equation}
and
\begin{eqnarray}\label{eq:R}
\ef_k^{\text{R}}(z)&=&\sqrt{\frac{2}{D\gamma_{kD}}}\,\frac{\sin(kz)+(k/c_1)\cos(k z)}{\sqrt{1+k^2/c_1^2}},\;k=k_\nu,\nonumber\\ &&\nu=1,2,\dots,\infty.
\end{eqnarray}
\end{subequations}
In the latter case of RBCs, the discrete values $k_\nu\equiv \mathsf{k}_\nu/D$ follow from the BC~\eqref{eq:BCR} at ${z=D}$. Here, the dimensionless $\mathsf{k}_\nu$ are given by the zeros of the function \cite{SD08,DS11,DrFSchmidt,fn3}
\begin{equation}\label{eq:Rdef}
R_{\scc_1,\scc_2}(\mathsf{k})=(\scc_1\scc_2/\sk-\sk)\sin(\sk)+(\scc_1+\scc_2)\cos(\sk),
\end{equation}
where $\scc_j=c_jD$, $j=1,2$.
Further, $\gamma_{\sk=kD}$ denotes the normalization factor
\begin{equation}
\gamma_{\sk}=1+\frac{\scc_1}{\scc_1^2+\sk^2}+\frac{\scc_1}{\scc_1^2+\sk^2},
\end{equation}
chosen such that the eigenfunctions are orthonormalized.

The BCs of the above eigenfunctions satisfy, e.g., 
\begin{align}
\ef_k^{\text{P}}(z) &= \ef_k^{\text{P}}(z+D), &\ef_k^{\text{A}}(z)&=-\ef_k^{\text{A}}(z+D),\nonumber\\ \ef_k^{\text{DD}}(0)&=\ef_k^{\text{DD}}(D)=0, &\ef_k^{\text{DN}}(0)&=\partial_z\ef_k^{\text{DN}}(D)=0,\nonumber\\
\end{align}
and Eq.~\eqref{eq:BCR}, carry over to the field operators $\psih(\bm{x})$ and $\psih^\dagger(\bm{x})$. The commutation relations of the latter are
\begin{eqnarray}
[\psih(\bm{x}),\psih^\dagger(\bm{x}')]&=&\delta(\bm{x}-\bm{x}'),\nonumber \\{}
[\psih(\bm{x}),\psih(\bm{x}')]&=&[\psih^\dagger(\bm{x}),\psih^\dagger(\bm{x}')]=0.
\end{eqnarray}

Following common practice \cite{WRFS86,BDZ08}, we  take the pair potential $u(\bm{x})$   to be short ranged and of the form
\begin{equation}\label{eq:u0d}
u(\bm{x})=\ub\,\delta(\bm{x}),\;\;\;\ub=\frac{4\pi^{d/2}\hbar^2a^{d-2}_s}{\Gamma(d/2-1)\,m},
\end{equation}
where $a_s$ is the $s$-wave scattering length \cite{WRFS86}.

We will be concerned with three different Bose~gas models on a film $[0,L]^{d-1}\times[0,D]$: the ideal Bose gas, the imperfect Bose gas model   investigated in \cite{NJN13}, \cite{JN13}, and \cite{NP11}, and the interacting Bose gas with $n$ internal degrees of freedom in the limit $n\to\infty$. In the case of the ideal Bose gas, the interaction $u(\bm{x})$ is zero. The imperfect Bose gas model results from the above specified interacting one if the potential energy term $U[\psih]$ is approximated by $U_{\text{impBG}}=a\hat{N}^2/2V$, where  
\begin{equation}
\hat{N}=\int_{\mathfrak{V}}d^dx\,\psih^\dagger(\bm{x})\psih(\bm{x})
\end{equation}
is the number operator and $V=AD$ means the hypervolume of the system. Thus the Hamiltonian of the imperfect Bose gas reads as
\begin{equation}\label{eq:HIBG}
\hat{H}_{\text{ImpBG}}=T[\psih^\dagger,\psih]+a\frac{\hat{N}^2}{2V}.
\end{equation}

This model looks unphysical in that each of the $N(N-1)/2=N^2/2+\Or(N)$ pairs of bosons gives the same contribution to $U_{\text{impBG}}$ independent of the  separation of the two bosons of each pair. It can be obtained by taking the limit $\gamma\to 0$ of a repulsive integrable Kac-type pair potential such as $u_\gamma(x)=\gamma^d \,\rme^{-\gamma x}$ whose strength and inverse range are both controlled by the same parameter $\gamma>0$ \cite{JN13,NJN13,NP11}. This  is analogous to the well-known rigorous derivation of the van der Waals theory for a classical gas of particles interacting through an attractive Kac pair potential and a repulsive hard core \cite{KUH63}, and hence reveals the mean-field nature of the approximation to which the model corresponds. The mentioned equivalence of the imperfect Bose gas with an interacting Bose gas with $n$ internal degrees of freedom in the limit ${n\to\infty}$ we are going to prove below provides an even nicer justification of the former model because the latter involves a physically reasonable short-ranged pair potential. 

To define the interacting Bose gas with $n$ internal degrees of freedom (called interacting $n$-component Bose gas model), we replace $\psih$ by an $n$-component operator $\bm{\psih}=(\psih_\alpha)$ with $\alpha=1,\dotsc,n$ and consider the Hamiltonian
\begin{equation}\label{eq:Hamn}
\hat{H}_n=T[\hat{\bm{\psi}}^\dagger,\hat{\bm{\psi}}]+\frac{\ub}{2n}\int\rmd^dx\,\psih^\dagger_\alpha(\bm{x})\psih^\dagger_\beta(\bm{x})\, \psih_\beta(\bm{x})\psih_\alpha(\bm{x}),
\end{equation}
where pairs of internal  indices $\alpha,\beta$ are to be summed from $1$ to $n$.  It will become clear below that  the imperfect Bose gas  is equivalent to this model in the limit $n\to\infty$.

Let $\Xi(T,\mu,D,L)$ denote the grand partition function
\begin{equation}
\Xi_(T,\mu,D,L)=\Tr\,\rme^{-\beta(\hat{H}-\mu\hat{N})}
\end{equation}
of any of these models with Hamiltonian $\hat{H}$. With a view to our subsequent analysis it will be helpful to recall its coherent-state path-integral  representation. 
To this end, we introduce complex-valued  fields $\bm{\psi}(\bm{x},\tau)=(\psi_\alpha(\bm{x},\tau))$ and $\bm{\psi}^*(\bm{x},\tau)=(\psi^*_\alpha(\bm{x},\tau))$  satisfying PBCs $\bm{\psi}(\bm{x},\tau)=\bm{\psi}(\bm{x},\tau+\beta\hbar)$ and $\bm{\psi}^*(\bm{x},\tau)$ $=\bm{\psi}^*(\bm{x},\tau+\beta\hbar)$ in imaginary time $\tau$. Owing to these BCs,  the Bose fields   $\bm{\psi}(\bm{x},\tau)$ and $\bm{\psi}^*(\bm{x},\tau)$ can be decomposed into Fourier series (see, e.g., \cite{Tau14})
\begin{equation}\label{eq:FTpsi}
\psi_\alpha(\bm{x},\tau)=\frac{1}{\beta\hbar}
\sum_{\rho\in\mathbb{Z}}\psi_{\alpha,\rho}(\bm{x})\,\rme^{-\rmi\omega_\rho\tau},\quad \omega_\rho=\frac{2\pi}{\beta\hbar}\rho,
\end{equation}
involving the bosonic Matsubara frequencies $\omega_\rho$ and the Fourier coefficients
\begin{equation}
\psi_{\alpha,\rho}(\bm{x})=\int_0^{\beta\hbar}\rmd{\tau}\,\psi_\alpha(\bm{x},\tau)\,\rme^{\rmi\omega_\rho\tau}.
\end{equation}
For later use,  we split $\psi_\alpha(\bm{x},\tau)$ into its $\tau$-independent $({\rho=0})$~part $\Psi_\alpha(\bm{x}) \equiv (\beta\hbar)^{-1} \psi_{\alpha,0}(\bm{x})$ and a remainder $\Psi^{\ne }_\alpha(\bm{x},\tau)$, writing \cite{fn10,BBZ-J00,Kas04}
\begin{equation}
\psi_{\alpha}(\bm{x},\tau)=\Psi_{\alpha}(\bm{x})+\Psi^{\ne }_\alpha(\bm{x},\tau).
\end{equation}

All BCs along the $z$~direction considered above for the operators $\psih_\alpha(\bm{y},z)$ and $\psih^*_\alpha(\bm{y},z)$  carry over to the Bose fields $\psi_\alpha(\bm{y},z,\tau)$ and $\psi^*_\alpha(\bm{y},z,\tau)$.

The coherent-state path-integral representation of the grand partition function reads as
 \begin{equation}\label{eq:Xi}
\Xi(T,\mu,D,L)=\int\mathcal{D}[\bm{\psi}^*,\bm{\psi}]\,\rme^{-S[\bm{\psi}^*,\bm{\psi}]/\hbar},
\end{equation}
where the action $S[\bm{\psi}^*,\bm{\psi}]$, in the case of the interacting Bose gas with Hamiltonian $\hat{H}_n$, is given by
\begin{subequations}\label{eq:S}
\begin{equation}\label{eq:Ssum}
S[\bm{\psi}^*,\bm{\psi}]=S_0[\bm{\psi}^*,\bm{\psi}]+ S_{\text{int}}[\bm{\psi}^*,\bm{\psi}]
\end{equation}
with
\begin{eqnarray}\label{eq:S0}
S_0[\bm{\psi}^*,\bm{\psi}]&=&\int_0^{\beta\hbar}\rmd\tau\int_{\mathfrak{V}}\rmd^dx\bigg[\psi_\alpha^*(\bm{x},\tau)\bigg(\hbar\partial_\tau\nonumber\\ 
&&\strut -\frac{\hbar^2}{2m}\nabla^2-\mu\bigg)\psi_\alpha(\bm{x},\tau)\bigg]
\end{eqnarray}
and
\begin{eqnarray}\label{eq:Sint}
S_{\text{int}}[\bm{\psi}^*,\bm{\psi}]&=&\frac{\ub}{2n}\int_0^{\beta\hbar}\rmd\tau\int_{\mathfrak{V}}\rmd^dx\big[\psi_\alpha^*(\bm{x},\tau)\psi_\beta^*(\bm{x},\tau)\nonumber\\&&\times \psi_\beta(\bm{x},\tau)\psi_\alpha(\bm{x},\tau)\big].
\end{eqnarray}
\end{subequations}

Near the bulk critical temperature, the part of the action involving Matsubara frequencies $\omega_\rho\ne 0$ is expected to give only exponentially small corrections. In fact, we will verify below that these corrections are down by factors $\rme^{-2\pi\sqrt{2}\,D/\lamth}$ at the bulk critical point  for PBCs and ABCs,  but smaller by its square $\rme^{-4\pi\sqrt{2}\,D/\lamth}$ for free BCs such as DDBCs, DNBCs, NNBCs, and RBCs. Ignoring these $\omega_\rho\ne0$ parts of the action by making the  replacement $\bm{\psi}\to\bm{\Psi}$, we arrive at the ``classical'' action $S[\bm{\Psi}^*,\bm{\Psi}]$, which can be conveniently written in terms of a rescaled real-valued $(2n)$-component order-parameter field $\bm{\phi}=(\phi_{\alpha_2})_{\alpha_2=1}^{2n}$ formed from the real and imaginary parts of $\Psi_\alpha$ such that
 \begin{equation}\label{eq:clresc}
\Psi_\alpha(\bm{x})=\frac{\sqrt{2\pi}}{\lamth}\,[\phi_{2\alpha-1}(\bm{x})+\rmi\phi_{2\alpha}(\bm{x})],\;\;\alpha=1,\dotsc,n.
\end{equation}
One finds
\begin{align}\label{eq:Scl}
S^{\text{cl}}[\bm{\phi}]&\equiv S[\bm{\Psi}^*,\bm{\Psi}]\nonumber \\
&=\int_{\mathfrak{V}}\rmd^dx\bigg[\frac{1}{2}\bm{\phi}\cdot(-\nabla^2\bm{\phi})+\frac{1}{2}\xi_{\text{id}}^{-2}\phi^2+\frac{g/n}{4!}\,\phi^4\bigg] \quad
\end{align}
with
\begin{equation}\label{eq:gdef}
g=\frac{12 m^2\ub}{\beta\hbar^4}
=\frac{96\,a_s^{d-2}\pi^{d/2+1}}{\lamth^2\Gamma(d/2-1)}.
\end{equation}
Here $\phi=\big(\sum_{\alpha_2=1}^{2n}\phi_{\alpha_2}^{2}\big)^{1/2}$, $(\nabla\bm{\phi})^2$ stands for $\sum_{\alpha_2=1}^{2n}(\nabla\phi_{\alpha_2})^2$, and we have added the subscript $\text{id}$ to emphasize that $\xi_{\text{id}}$ is the ideal Bose gas quantity defined in Eq.~\eqref{eq:xiid}.

\section{Scaling functions of the ideal Bose gas\label{eq:scfcideal}}
\subsection{The cases of  periodic, antiperiodic, Dirichlet-Dirichlet, Dirichlet-Neumann, and Neumann-Neumann boundary conditions}

For the case of the ideal  Bose gas (with ${n=1}$ components) the scaling functions  $\Upsilon_d^{\text{P}}$, $\Upsilon_d^{\text{DD}}$, and $\Upsilon_d^{\text{NN}}$ can be gleaned from \cite{MZ06}. Since their $n>1$ analogs follow upon multiplication by $n$, we set $n=1$ unless stated otherwise (see Sec.~\ref{sec:intBG}).
In Appendix~\ref{app:idBG} we present a slightly different calculation of $\Upsilon_d^{\text{P}}$, determine $\Upsilon_d^{\text{A}}$, and recapitulate how $\Upsilon_d^{\text{DD}}$ and $\Upsilon_d^{\text{NN}}$ can be computed. The results read as
\begin{eqnarray} \label{eq:Ydperres}
\lefteqn{\Upsilon_d^{\text{P}}(x_\lambda,x_\xi)}&&\nonumber\\
&=&-2x_\lambda^d\sum_{s=1}^\infty s^{-d/2-1}\sum_{j=1}^\infty\rme^{-\pi j^2 x_\lambda^2/s-sx_\xi^2/4\pi x_\lambda^2}\nonumber\\
&=&-x_\lambda^d\sum_{s=1}^\infty s^{-d/2-1}\rme^{-sx_\xi^2/4\pi x_\lambda^2}\big[\vartheta_3\big(0,\rme^{-\pi x_\lambda^2/s}\big)-1\big],\qquad
\end{eqnarray}
\begin{eqnarray} \label{eq:Ydapres}
\lefteqn{\Upsilon_d^{\text{A}}(x_\lambda,x_\xi)}&&\nonumber\\
&=&-2x_\lambda^d\sum_{s=1}^\infty s^{-d/2-1}\sum_{j=1}^\infty(-1)^j\rme^{-\pi j^2 x_\lambda^2/s-sx_\xi^2/4\pi x_\lambda^2}\nonumber\\
&=&-x_\lambda^d\sum_{s=1}^\infty s^{-d/2-1}\rme^{-sx_\xi^2/4\pi x_\lambda^2}\big[\vartheta_4\big(0,\rme^{-\pi x_\lambda^2/s}\big)-1\big],\qquad
\end{eqnarray}
and
\begin{eqnarray} \label{eq:UpsDDNN}
\lefteqn{\Upsilon_d^{\text{DD}}(x_\lambda,x_\xi)=\Upsilon_d^{\text{NN}}(x_\lambda,x_\xi)}&&\nonumber\\
&=&-2x_\lambda^d\sum_{s=1}^\infty s^{-d/2-1}\sum_{j=1}^\infty\rme^{-4\pi j^2 x_\lambda^2/s-sx_\xi^2/4\pi x_\lambda^2}\nonumber\\
&=&-x_\lambda^d\sum_{s=1}^\infty s^{-d/2-1}\rme^{-sx_\xi^2/4\pi x_\lambda^2}\big[\vartheta_3\big(0,\rme^{-4\pi x_\lambda^2/s}\big)-1\big],\qquad
\end{eqnarray}
where $\vartheta_3(z,q)=\sum_{j=-\infty}^\infty q^{j^2}\rme^{2\pi\rmi z}$ and $\vartheta_4(z,q)=\sum_{j=-\infty}^\infty (-1)^jq^{j^2}\rme^{2\pi\rmi z}$ are Jacobi theta functions \cite{NIST:DLMF,Olver:2010:NHMF}.

In the limit $x_\lambda\to\infty$, the sums $\sum_{s=1}^\infty\ldots $ become Riemann sums for integrals $\int_1^\infty\rmd{s}\ldots $. Furthermore, the lower integration limit can be changed from $1$ to $0$ since the differences $\int_0^1\rmd{s}\ldots$ vanish in the limit  $x_\lambda\to 0$. Performing the integrals yields
\begin{subequations}
\begin{eqnarray}
\lefteqn{\Theta^{\text{P}}_d(x_\xi)}&&\nonumber\\ &=&-4(2\pi)^{-d/2}\,x_\xi^{d/2}\sum_{j=1}^\infty j^{-d/2}\,K_{d/2}(jx_\xi)\label{eq:Thetadpera}\\
&=&-\frac{2\,K_{d-1}}{d-1}\,x_\xi^d\int_1^\infty\rmd{t}\,\frac{(t^2-1)^{(d-1)/2}}{\rme^{tx_\xi}-1}, \label{eq:Thetadperb}
\end{eqnarray}
\end{subequations}
 \begin{subequations}
\begin{eqnarray}
\lefteqn{\Theta^{\text{A}}_d(x_\xi)}&&\nonumber\\ &=&4(2\pi)^{-d/2}\,\,x_\xi^{d/2}\sum_{j=1}^\infty \frac{(-1)^{j+1}}{ j^{d/2}}\,K_{d/2}(jx_\xi)\label{eq:Thetadapa} \quad\\
&=&\frac{2\,K_{d-1}}{d-1}\,x_\xi^d\int_1^\infty\rmd{t}\,\frac{(t^2-1)^{(d-1)/2}}{\rme^{tx_\xi}+1}.\label{eq:Thetadapb}, 
\end{eqnarray}
\end{subequations}
and
\begin{subequations}\label{eq:ThetadDD}
\begin{eqnarray}
\lefteqn{\Theta^{\text{DD}}_d(x_\xi)=\Theta^{\text{NN}}_d(x_\xi)}&& \nonumber\\ &=&\strut -4(2\pi)^{-d/2}\,\,x_\xi^{d/2}\sum_{j=1}^\infty j^{-d/2}\,K_{d/2}(2jx_\xi),\label{eq:ThetadDDa}\\
&=&\strut-\frac{2\,K_{d-1}}{d-1}\,x_\xi^d\int_1^\infty\rmd{t}\,\frac{(t^2-1)^{(d-1)/2}}{\rme^{2tx_\xi}-1},\label{eq:ThetadDDb}
\end{eqnarray}
\end{subequations}
where 
\begin{equation}\label{eq:Kddef}
K_d\equiv\frac{2}{(4\pi)^{d/2}\Gamma(d/2)},
\end{equation}
while $K_{d/2}(x)$ is a modified Bessel function of the second kind.

For DNBCs, the corresponding result reads as
 \begin{subequations}\label{eq:ThetadDN}
\begin{eqnarray}
\lefteqn{\Theta^{\text{DN}}_d(x_\xi)}&&\nonumber \\  &=&4(2\pi)^{-d/2}\,x_\xi^{d/2}\sum_{j=1}^\infty \frac{(-1)^{j+1}}{ j^{d/2}}\,K_{d/2}(2jx_\xi)\label{eq:ThetadDNa}\quad\\
&=&\frac{2\,K_{d-1}}{d-1}\,x_\xi^d\int_1^\infty\rmd{t}\,\frac{(t^2-1)^{(d-1)/2}}{\rme^{2tx_\xi}+1}.\label{eq:ThetadDNb}
\end{eqnarray}
\end{subequations}
This result most easily follows by setting $(\scc_1,\scc_2)=(\infty,0)$ in the result for RBCs  derived in the next subsection [see Eqs.~\eqref{eq:UpsRresd} and \eqref{eq:ThetaR}] and integrating by parts.

The integral forms given in Eqs.~\eqref{eq:Thetadperb}--\eqref{eq:ThetadDNb} follow from Eq.~(6.8) of \cite{KD92a} upon setting $n=2$ there. To verify that the series \eqref{eq:Thetadpera}--\eqref{eq:ThetadDNa} can be summed in this manner, one can substitute 
the expansion
\begin{eqnarray}
 [\rme^{tx}\mp1]^{-1}&=&\sum_{j=1}^\infty (\pm1)^{j+1}\,\rme^{-jtx}
 \end{eqnarray}
 with $x=x_\xi$ or $2x_\xi$ into these integrals and integrate termwise.
 
It should be obvious that the above results for $\Theta^{\text{BC}}_d(x_\xi)$ with $\text{BC}=\text{P},\text{A},\text{DD},\text{NN}$, and $\text{DN}$ are identical to  those that follow from the ${g=0}$ analog of $S^{\text{cl}}[\bm{\phi}]$. 

\begin{figure}[htbp]
\begin{center}
\includegraphics[width=0.95\columnwidth]{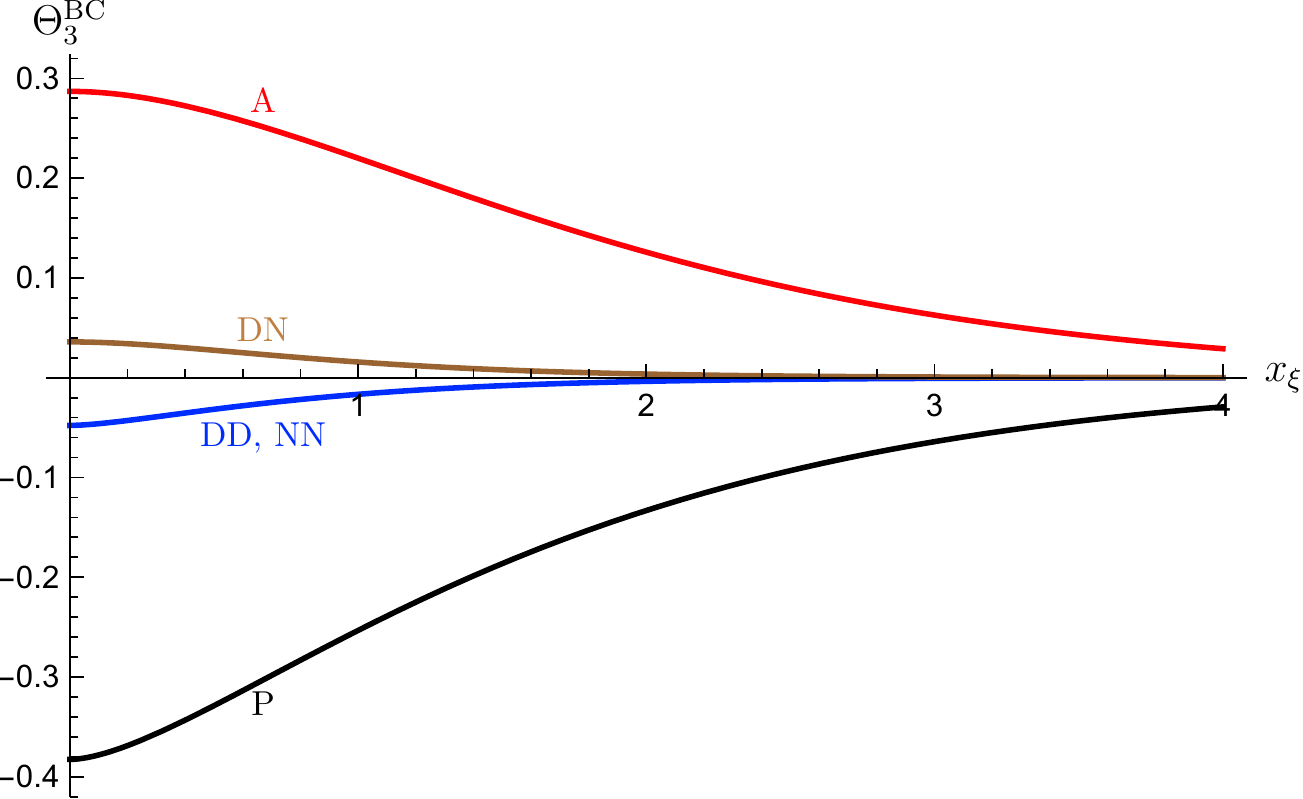}
\caption{Classical scaling functions $\Theta^{\text{BC}}_3(x_\xi)$ for $n=1$ and $\text{BC}=\text{P},\text{A},\text{DD},\text{NN}$, and $\text{DN}$. Their values at $x_\xi$ are  the corresponding Casimir amplitudes $\Delta_d^{\text{BC}}$.}
\label{fig:fig1DR16b}
\end{center}
\end{figure}

When $d=3$, the results for $\Theta^{\text{BC}}_d$ given in Eqs.~\eqref{eq:Thetadpera}--\eqref{eq:ThetadDNb} can be expressed in closed form in terms of polylogarithms \cite{NIST:DLMF,Olver:2010:NHMF}; one has
\begin{subequations}
\begin{eqnarray}
\Theta_3^{\text{P}}(x_\xi)&=&-\frac{1}{\pi}\,[\Li_3(\rme^{-x_\xi})+x_\xi\,\Li_2(\rme^{-x_\xi})],\label{eq:Theta3per}\\
\Theta_3^{\text{A}}(x_\xi)&=&-\frac{1}{\pi}\,[\Li_3(-\rme^{-x_\xi})+x_\xi\,\Li_2(-\rme^{-x_\xi})],\label{eq:Theta3ap}\\
\Theta_3^{\text{DD}}(x_\xi)&=&\Theta_3^{\text{NN}}(x_\xi)\nonumber\\ &=&\strut-\frac{1}{8\pi}\,[\Li_3(\rme^{-2x_\xi})+2x_\xi\,\Li_2(\rme^{-2x_\xi})],\qquad\;\;\label{eq:Theta3DD} \\
\Theta_3^{\text{DN}}(x_\xi)&=&-\frac{1}{8\pi}\,[\Li_3(-\rme^{-2x_\xi})+2x_\xi\,\Li_2(-\rme^{-2x_\xi})].\quad\nonumber\\ \label{eq:Theta3DN}
\end{eqnarray}\end{subequations}
These functions are plotted in Fig.~\ref{fig:fig1DR16b}.

\begin{figure}[htbp]
\begin{center}
\includegraphics[width=0.95\columnwidth]{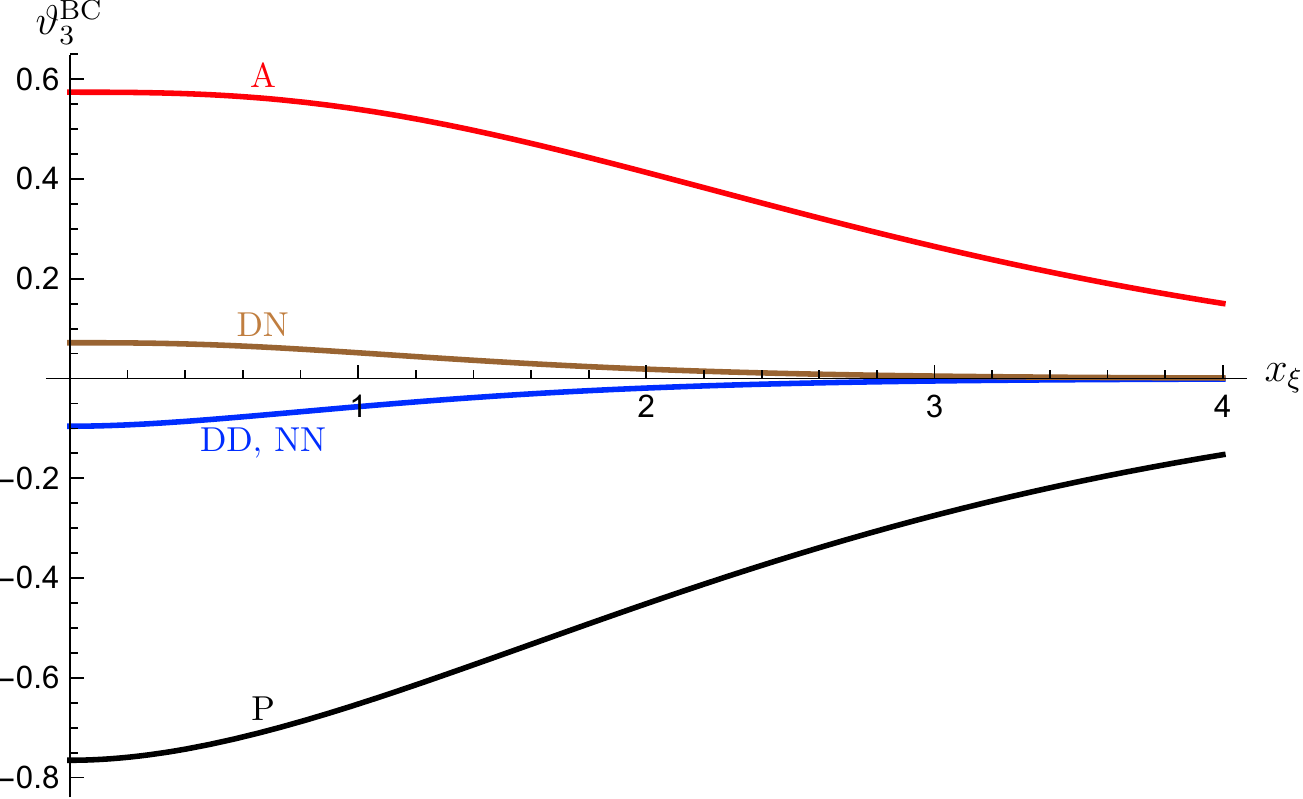}
\caption{Classical scaling functions $\vartheta^{\text{BC}}_3(x_\xi)$ for $n=1$ and $\text{BC}=\text{P},\text{A},\text{DD},\text{NN}$, and $\text{DN}$.}
\label{fig:fig2DR16b}
\end{center}
\end{figure}

The corresponding results for the scaling functions of the Casimir force follow in a straightforward fashion via
\begin{equation}\label{eq:Thetavarthetarel}
 \vartheta_{d}^{\text{BC}}(x_\xi)=(d-1)\,\Theta_{d}^{\text{BC}}(x_\xi)-x_\xi\frac{\rmd}{\rmd x_\xi}\Theta_{d}^{\text{BC}}(x_\xi).
 \end{equation}
They read as
\begin{subequations}
\begin{eqnarray}
\vartheta_3^{\text{P}}(x_\xi)&=&\frac{1}{\pi}\big[x_\xi^2\ln\big(1-\rme^{-x_\xi}\big)-2\,\Li_3\big(-\rme^{-x_\xi}\big)\nonumber\\&&\strut -2x_\xi\,\Li_2\big(\rme^{-x_\xi}\big)\big],\label{eq:vartheta3per}\\
\vartheta_3^{\text{A}}(x_\xi)&=&\frac{1}{\pi}\big[x_\xi^2\ln\big(1+\rme^{-x_\xi}\big)-2\,\Li_3\big(\rme^{-x_\xi}\big)\nonumber\\&&\strut -2x_\xi\,\Li_2\big(-\rme^{-x_\xi}\big)\big],\label{eq:vartheta3ap}\\
\vartheta_3^{\text{DD}}(x_\xi)&=&\frac{1}{4\pi}\big[x_\xi^2\ln\big(1-\rme^{-2x_\xi}\big)-\Li_3\big(\rme^{-x_\xi}\big)\nonumber\\&&\strut -2x_\xi\,\Li_2\big(\rme^{-2x_\xi}\big)\big],\label{eq:vartheta3DD} \\
\vartheta_3^{\text{DN}}(x_\xi)&=&\frac{1}{4\pi}\big[2x_\xi^2\ln\big(1+\rme^{-2x_\xi}\big)-\Li_3\big(-\rme^{-x_\xi}\big)\nonumber\\&&\strut -2x_\xi\,\Li_2\big(-\rme^{-2x_\xi}\big)\big], \label{eq:vartheta3DN}
\end{eqnarray}\end{subequations}
and are plotted in Fig.~\ref{fig:fig2DR16b}.

The Casimir amplitudes $\Delta^{\text{BC}}_d=\Theta^{\text{BC}}_d(0)$ that follow from the above Eqs.~\eqref{eq:Thetadpera}--\eqref{eq:ThetadDNb} and \eqref{eq:Theta3per}--\eqref{eq:Theta3DN} for the scaling functions, 
\begin{subequations}
\begin{eqnarray}
\Delta^{\text{P}}_d&=&-2\pi^{-d/2}\Gamma(d/2)\,\zeta(d)\nonumber\\ &\mathop{=}\limits_{d=3}&-\zeta(3)/\pi,\\
\Delta^{\text{A}}_d&=&2(1-2^{1-d})\pi^{-d/2}\,\Gamma(d/2)\,\zeta(d)\nonumber\\ &\mathop{=}\limits_{d=3}&3\,\zeta(3)/4\pi,
\end{eqnarray}
\begin{eqnarray}
\Delta^{\text{DD}}_d=\Delta^{\text{NN}}_d&=&-2^{1-d}\pi^{-d/2}\Gamma[d/2]\,\zeta(d)\nonumber \\ &\mathop{=}\limits_{d=3}&-\zeta(3)/8\pi,
\end{eqnarray}
and
\begin{eqnarray}
\Delta^{\text{DN}}_d&=&(1-2^{1-d})2^{1-d}\pi^{-d/2}\Gamma(d/2)\zeta(d)\nonumber\\ &\mathop{=}\limits_{d=3}&3\,\zeta(3)/32\pi,
\end{eqnarray}
\end{subequations}
agree with the familiar one-loop results for the $O(2)$ massless $\phi^4$ theory (cf.\ Eq.~(5.7) of \cite{KD92a} and Eq.~(4.4) of \cite{Sym81}).

\subsection{Case of  Robin boundary conditions\label{sec:BCR}}

We now turn to the case of Robin boundary conditions with $c_1 \ge 0$ and $c_2\ge 0$. We assume that $c_1+c_2>0$, so that not both variables vanish simultaneously. In this case, the discrete $k$~values $k_\nu$, $\nu=1,2,\ldots,\infty$, introduced in  Eq.~\eqref{eq:R} are all positive. 
It is convenient to introduce the density of states
\begin{equation}\label{eq:rhodef}
\rho(E;D,c_1,c_2)=\sum_{\nu=1}^\infty \delta(E-k_\nu^2),\quad E>0,
\end{equation}
which can be expressed as
\begin{equation}\label{eq:rhoRandT}
\rho(E;D,c_1,c_2)=-\frac{1}{\pi}\Im\,T(E+\rmi 0)\equiv -\frac{1}{\pi}\Im\Tr G(E+\rmi 0)
\end{equation}
in terms of the trace $T(z)\equiv T(z;D,c_1,c_2)$ of the Green's function
\begin{equation}\label{eq:GR}
G(z)\equiv G(z;D,c_1,c_2)=\big(z+\partial_z^2\big)^{-1},
\end{equation}
where the operator $\partial_z^2$ is subject to the RBCs specified in Eq.~\eqref{eq:BCR}.

With the aid of the generalized function $\rho(E;D,c_1,c_2)$ the reduced grand potential $\varphi_d^{\text{R}}$ can be written as
\begin{align}\label{eq:varphiitrho}
&\varphi^{\text{R}}_d(T,\mu,D,c_1,c_2)\nonumber\\ &=-\lamth^{1-d}\int_0^\infty \rmd{E}\,\Li_{\frac{d+1}{2}}\Big(\rme^{\beta\mu -\lamth^2 E/4\pi}\Big)\rho(E;D,c_1,c_2).
\end{align}

The function $T(z;D,c_1,c_2)$ is given by
\begin{equation}
T(z;D,c_1,c_2)=\sum_{\nu=1}^\infty \big(z-k_\nu^2\big)^{-1},\quad z\in\mathbb{C}\setminus[0,\infty),
\end{equation}
and has the scaling property
\begin{eqnarray}
T(z;D,c_1,c_2)=D^2\,T(D^2z;1,Dc_1,Dc_2).
\end{eqnarray}
Recalling that  the scaled momenta $\sk_\nu=Dk_\nu$ are the zeros of the function  $R_{\scc_1,\scc_2}(\sk)$ and that this function is even in $\sk$, one concludes that $\partial_\zeta\ln R_{\scc_1,\scc_2}(\sqrt{\zeta})$ has simple poles at $\zeta=\sk_\nu^2$ with residues $1$ and therefore agrees with $T(\zeta;1,\scc_1,\scc_2)$. Hence,
\begin{widetext}
 \begin{equation}\label{eq:Tscaled}
T(\zeta;1,\scc_1,\scc_2)=-\frac{1}{2\zeta}\frac{(\zeta-\scc_1\scc_2)\sqrt{\zeta}\cos\sqrt{\zeta}+[\scc_1\scc_2+(1+\scc_1+\scc_2)\zeta]\sin\sqrt{\zeta}}{(\scc_1+\scc_2)\sqrt{\zeta}\cos\sqrt{\zeta}+(\scc_1\scc_2-\zeta)\sin\sqrt{\zeta}}\,,
 \end{equation}
 where the branch cut of $\sqrt{\zeta}$ is taken along the positive real axis.
\end{widetext}
Choosing  $\Im\,\zeta>0$, we can expand the result  in powers of $D$ about $D=\infty$.We thus obtain the decomposition
\begin{equation}\label{eq:Tdec}
T(z;D,c_1,c_2)=D\,T_{\text{b}}(z)+\sum_{j=1}^2T_{\text{s}}(z;c_j)+T_{\text{res}}(z;D,c_1,c_2)
\end{equation}
into bulk, surface, and residual contributions with
\begin{eqnarray}\label{eq:Tbdef}
T_{\text{b}}(z;D,c_1,c_2)&\equiv&\lim_{D\to\infty}D^{-1}T(z;D,c_1,c_2)\nonumber\\ &=&\frac{-\rmi}{2\sqrt{z}}
\end{eqnarray}
and 
\begin{eqnarray}\label{eq:Ts}
T_{\text{s}}(z;c)&\equiv&\lim_{D\to\infty}[T(z;D,c,c)-D\,T_{\text{b}}(z)]/2,\nonumber\\ &=&
\frac{1}{4z}\,\frac{\sqrt{z}-\rmi c}{\sqrt{z}+\rmi c}.
\end{eqnarray}
The reader may want to check that the result~\eqref{eq:Tbdef} is consistent with what one gets from Eqs~\eqref{eq:rhoRandT} and \eqref{eq:GR} using the fact that  $T_{\text{b}}(z)$ is independent of the BC along with  $\lim_{D\to\infty}D^{-1}\Tr f(-\partial_z^2)=\int_{-\infty}^\infty\frac{\rmd{k}}{2\pi}f(k^2)$. The associated bulk density of states becomes
\begin{equation}\label{eq:rhob}
\rho_{\text{b}}(E)=\frac{1}{2\pi\sqrt{E}}.
\end{equation}
One can easily check that Eq.~\eqref{eq:varphib} for $\varphi_{\text{b},d}$ is recovered from Eq.~\eqref{eq:varphiitrho} upon substituting $\rho(E)$ by the foregoing result for $\rho_{\text{b}}(E)$.

In order to determine the surface quantity 
\begin{equation}
\rho_{\text{s}}(E;c)=-\frac{1}{\pi}\Im\, T_{\text{s}}(E+\rmi 0;c)
\end{equation}
from Eq.~\eqref{eq:Ts}, the familiar Sokhatsky-Weierstra{\ss} identity
\begin{equation}\label{eq:SWI}
\frac{1}{E+\rmi 0}=\frac{P}{E}-\rmi\pi\,\delta(E)
\end{equation}
is needed, where $P$ denotes the principal value. One finds
\begin{equation}\label{eq:rhos}
\rho_{\text{s}}(E;c)
=\frac{\theta(E)}{2\pi\sqrt{E}}\,\frac{c}{c^2+E}-\frac{1}{4}\,\delta(E),
\end{equation}
where we have included the Heaviside function $\theta(E)$ in the first term, which originated from the principal-value term on the right-hand side of Eq.~\eqref{eq:SWI}, to emphasize that this distribution is integrated only over the positive real axis. By contrast,  the $E$~integration involving $\delta(E)$ is to be extended over the full real axis, so that its action on a test function $g(E)$ yields the usual result $g(0)$. 

The result~\eqref{eq:rhos}  can be substituted  into Eq.~\eqref{eq:varphiitrho} to determine $\varphi_{\text{s},d}$. Upon transforming to the integration variable $k=\sqrt{E}$ and exploiting the evenness of the integrand in $k$, one arrives at
\begin{align}\label{eq:varphisRrho}
&\varphi^{\text{R}}_{\text{s},d}(T,\mu,c)\nonumber\\
&=\lamth^{1-d}\bigg[\frac{1}{4}\,\Li_{\frac{d+1}{2}}(\rme^{\beta\mu})\nonumber\\ &\strut
 -\int_{-\infty}^\infty\frac{\rmd{k}}{2\pi}\Li_{\frac{d+1}{2}}\Big(\rme^{\beta\mu-\lamth^2k^2/4\pi}\Big)\,\frac{c}{c^2+k^2}\bigg].
\end{align}
In the limit ${c\to\infty}$ (corresponding to DBCs), the contribution from the integral vanishes. 
Thus
\begin{equation}\label{eq:varphisD}
\varphi^{\text{D}}_{\text{s},d}(T,\mu)\equiv\varphi^{\text{R}}_{\text{s},d}(T,\mu,\infty)=\frac{1}{4}\lamth^{1-d}\,\Li_{\frac{d+1}{2}}(\rme^{\beta\mu}).
\end{equation}
To determine $\varphi^{\text{R}}_{\text{s},d}(T,\mu,0)$ (corresponding to NBCs), one can use
\begin{equation}\label{eq:czerodelta}
\frac{c}{c^2+k^2}\xrightarrow[c\to 0]{}\pi\delta(k)
\end{equation}
to  find
\begin{equation}\label{eq:varphisN}
\varphi^{\text{N}}_{\text{s},d}(T,\mu)\equiv \varphi^{\text{R}}_{\text{s},d}(T,\mu,0)=-\frac{1}{4}\lamth^{1-d}\,\Li_{\frac{d+1}{2}}(\rme^{\beta\mu}).
\end{equation}
These results~\eqref{eq:varphisD} and \eqref{eq:varphisN} are consistent with the integral expressions given in Eq.~(11) of  \cite{MZ06} for the total surface contributions of $\varphi_{d=3}^{\text{BC}}$ for DDBCs  and NNBCs. They also imply that the surface contribution of $\varphi_d$ vanishes for DNBCs. 

At the Bose-Einstein transition in $d>2$ dimensions, the correlation length $\xi=\xi_{\text{id}}$ diverges ${\asprop (-\mu)^{-1/2}}$ as $\mu\to 0-$ and becomes much larger than the thermal de Broglie wavelength $\lamth({T\etwa \tc})$. If the approach to criticality occurs along a temperature path at fixed  density $\rho=-\partial\varphi_{\text{b},d}/\partial(\beta\mu)$, then $-\mu\asprop  \delta T^{2/(d-2)}$,  where 
\begin{equation}
\delta T=[T-T_{\text{c}}(\rho)]/T_{\text{c}}(\rho),
\end{equation} 
so that $\xi_{\text{id}}\sim \delta T^{-\nu_{\text{id}}}$ with $\nu_{\text{id}}=(d-2)^{-1}$.

The correlation-length exponent $\nu_{\text{id}}$ and the other critical exponents of the ideal Bose gas may be understood as Fisher-renormalized exponents \cite{Fis68,WRFS86} of the Gaussian model; i.e., the specific-heat, order-parameter, susceptibility, and correlation-length exponents $\alpha_{\text{id}}$, $\beta_{\text{id}}$, $\gamma_{\text{id}}$, and $\nu_{\text{id}}$, respectively, follow from their Gaussian counterparts $\alpha_{\text{G}}$,\ldots via the relations
\begin{eqnarray}
\nu_{\text{id}}&=&\frac{\nu_{\text{G}}}{1-\alpha_{\text{G}}}=\frac{1}{d-2}=\gamma_{\text{id}}/2,\;\;\nu_{\text{G}}=\frac{1}{2},\nonumber\\
\beta_{\text{id}}&=&\frac{\beta_{\text{G}}}{1-\alpha_{\text{G}}}=\frac{1}{2},\;\;\beta_{\text{G}}=\nu_{\text{G}}\frac{d-2}{2}=\frac{d-2}{4},\nonumber\\
\alpha_{\text{id}}&=&\frac{-\alpha_{\text{G}}}{1-\alpha_{\text{G}}}=\frac{d-4}{d-2},\;\,\alpha_{\text{G}}=2-\frac{d}{2}.
\end{eqnarray}

The asymptotic critical behavior is known to be purely classical. The bulk universality class is that of a Gaussian model for a two-component real-valued order parameter with a mass term $\xi_{\text{id}}^{-2}\asprop (-\mu)\to 0$, but the above-mentioned renormalization of the critical exponents due to the fixed-density constraint must be taken into account. The length $\lamth(T)$ should drop out from the asymptotic critical behavior in appropriately normalized quantities. Let us therefore determine the limiting behavior of $\varphi_{\text{s},d}^{\text{R}}$ for $\lamth\to 0$. One possibility is to use the small-$x$ expansion \cite{fn4}
\begin{eqnarray}\label{eq:Lias}
\lefteqn{\Li_{(d+1)/2}(1-x)}&&\nonumber\\ &=&\begin{cases}\zeta(\frac{d+1}{2})+\zeta(\frac{d-1}{2})\,x+\Gamma(\frac{1-d}{2})\,x^{\frac{d-1}{2}}+\ldots&\text{for }3\ne d,\nonumber \\\frac{1}{6}\,\pi^2-x(1-\ln x)+\ldots&\text{for }d=3.
\end{cases}\\
\end{eqnarray}
for the polylogarithms in Eq.~\eqref{eq:varphisRrho}. In order to benefit from dimensional regularization we insert the above expansion for general $d\in(2,4)$ into the integral in Eq.~\eqref{eq:varphisRrho}. This leads us to
\begin{equation}\label{eq:varphisRexp}
\varphi_{\text{s},d}^{\text{R}}(T,\mu,c)=\varphi_{\text{s},d}^{\text{R}}(T,0)+f_{\text{s},d}^{\text{R}}(\xi,c)+\text{o}(\lamth)
\end{equation}
with
\begin{equation}
\varphi_{\text{s},d}^{\text{R}}(T,0;c)=-\frac{1}{4}\lamth^{1-d}\,\zeta\Big(\frac{d+1}{2}\Big)
\end{equation}
and
\begin{align}\label{eq:fsRint}
f^{\text{R}}_{\text{s},d}(\xi,c)=&\Gamma\Big(\frac{1-d}{2}\Big)\bigg[\frac{1}{4}\,(4\pi\xi^2)^{(1-d)/2}\nonumber\\ &\strut 
-\int_0^\infty\frac{\rmd{k}}{\pi}\left(\frac{k^2+\xi^{-2}}{4\pi}\right)^{(d-1)/2}\frac{c}{k^2+c^2}\bigg],
\end{align}
where $\xi$ again means $\xi_{\text{id}}$ and the omitted $\text{o}(\lamth)$ terms contain quantum corrections.

The result requires two comments. First, the integral in Eq.~\eqref{eq:varphisRrho} converges in the ultraviolet (UV) because the $k$ integration is smoothly cut off at $k_\lambda\etwa 4\pi/\lamth$. However, since $\lamth$ drops out from the integral of the expansion term associated with $f^{\text{R}}_{\text{s},d}$, a  UV cutoff is no longer present in it. Convergence of this integral is ensured only for $d<2$. For $2<d<4$, it must be regularized either by reintroducing a UV cutoff or else dimensionally. We prefer to use dimensional regularization. Assuming that $d<2$, the integral in Eq.~\eqref{eq:fsRint} can be computed by means of  {\sc Mathematica} \cite{Mathematica11}. One obtains
 \begin{eqnarray}
\lefteqn{f^{\text{R}}_{\text{s},d}(\xi,c)=c^{d-1}f_{\text{s},d}(\xi c,1)}&&\nonumber\\ &=&\frac{1}{2}\xi^{1-d}(4\pi)^{\frac{1-d}{2}} \bigg\{\Gamma\Big(\frac{1-d}{2}\Big)\bigg[\frac{1}{2}
-\frac{(c^2 \xi
 ^2-1)^{(d-1)/2}}{\sin(\pi d/2)}\bigg]\nonumber\\ &&\strut+
\frac{\sqrt{\pi }  \, _2\tilde{F}_1[1/2,1;d/2+1;(c \xi)^{-2}]}{c \xi \sin(\pi d/2)}\bigg\},
\end{eqnarray}
where ${}_2\tilde{F}_1(a,b;c;z)$ is the regularized hypergeometric function $ {}_2{F}_1(a,b;c;z)/\Gamma(c)$. The result provides the analytic continuation to dimensions $d\ge 2$.

A second  necessary comment is that this expression does not reproduce the $c\to\infty$ limit of $f^{\text{R}}_{\text{s},d}$ given by the first term in Eq.~\eqref{eq:fsRint} (since the regularized integral vanishes for $c\to\infty$). The reason is that the sequence of the limits in which the lengths $1/c$ and $\lamth$ go to zero (or the cutoff $k_\lambda\to\infty$) do not commute. This noncommutabilty of the limits $c\to\infty$ of  the bare surface-enhancement variable $c$ and the UV momentum cutoff is well known from the classical theory \cite{Die86a}.

However, the result given in Eq.~\eqref{eq:fsRint} yields the correct limits $c\to 0$ with $\xi >0$ and $\xi\to\infty$ with $0<c<\infty$, namely,
\begin{equation}\label{eq:fsRc}
f^{\text{R}}_{\text{s},d}(\infty,c)=-K_{d-1}\frac{\pi}{d-1}\frac{c^{d-1}}{\sin(d\pi)}
\end{equation}
and
\begin{equation}\label{eq:fsRczero}
f^{\text{R}}_{\text{s},d}(\xi,0)=-\frac{K_{d-1}}{d-1}\frac{\pi}{4\cos(\pi d/2)}\,\xi^{1-d}.
\end{equation}
Equation~\eqref{eq:fsRc} complies with the result obtained in \cite{SD08}, \cite{DS11}, and \cite{DrFSchmidt}. Further,  Eq.~\eqref{eq:fsRczero} can  be  confirmed easily via Eq.~\eqref{eq:fsRint} using Eq.~\eqref{eq:czerodelta}.

The result~\eqref{eq:fsRint} can be checked by means of a purely classical calculation. Consider the free classical theory associated with the action~\eqref{eq:Scl} with ${g=0}$  for the semi-infinite system $z\ge 0$  and subject to the  RBC~\eqref{eq:BCR} with $c\equiv c_1$ at $z=0$. The Fourier transform of the corresponding free propagator $\langle \bm{\phi}(\bm{y},z)\cdot\bm{\phi}(\bm{0},z')\rangle/2n$ with respect to the $\bm{y}$~coordinate is given by
\begin{equation}\label{eq:gsipz}
\hat{G}[\kappa_p(\xi);z,z';c]=\frac{1}{2\kappa}\bigg[\rme^{-\kappa|z-z'|}-\frac{c-\kappa}{c+\kappa}\,\rme^{-\kappa(z+z')}\bigg]_{\kappa=\kappa_p(\xi)}
\end{equation}
where
\begin{equation}
\kappa_p(\xi)=\sqrt{p^2+\xi^{-2}}.
\end{equation}
Using this leads us to
\begin{eqnarray}
e_{\text{s},d}(\xi,c)&=&\pint\int_0^\infty\rmd{z}\Big[\hat{G}[\kappa_p(\xi);z,z;c]-(z=\infty)\Big]\nonumber\\
&=&\pint\bigg[\frac{1}{4\kappa^2}\frac{\kappa-c}{\kappa+c}\bigg]_{\kappa=\kappa_p(\xi)}
\end{eqnarray}
for the surface energy density, where we adopted the convenient notation
\begin{equation}
\pint=\int_{\mathbb{R}^{d-1}}\frac{\rmd^{d-1}p}{(2\pi)^{d-1}}.
\end{equation}
Integrating the result with respect to $\xi^{-2}$ yields
\begin{eqnarray}\label{eq:fs2}
f_{\text{s},d}(\xi,c)&=&c^{d-1}\pint\ln\frac{1+\kappa}{\sqrt{\kappa}}\bigg|_{\kappa=\kappa_p(\xi c)}\nonumber\\ &=&
c^{d-1}K_{d-1}\,J_d(\xi c)
\end{eqnarray}
with
\begin{eqnarray}\label{eq:Jd}
J_d( v)&=&\int_0^\infty\rmd{p}\,p^{d-2}\ln\bigg[\frac{1+\kappa_p(v)}{\sqrt{\kappa_p(v)}}\bigg]\nonumber\\
&=&\frac{1}{1-d}\int_0^\infty\frac{\rmd{p}\,p^d}{2\kappa^2_p(v)}\frac{\kappa_p(v)-1}{\kappa_p(v)+1},
\end{eqnarray}
where the second form follows upon integration by parts. The latter integral, which is UV divergent for dimensions $d>1$,  can be analytically continued  to $d>2$ in a straightforward manner. In the first form, $J_d(v)$  is  IR divergent for $d<1$ and UV divergent for $d>1$. To avoid the IR divergence, one can subtract from the integrand's logarithm its value at $p=0$, using the fact that $\int_0^\infty\rmd{p}\,p^{d-2}=0$ in dimensional regularization. The resulting integral thereby becomes well-defined for $d\in(-1,1)$ and can be analytically continued. 

Consistency with the results for $(\xi,c)=(\infty,c)$ and $(\xi,0)$ given in Eqs.~\eqref{eq:fsRc} and \eqref{eq:fsRczero} is easily checked by performing the required integrals. A proof that Eqs.~\eqref{eq:fs2} and \eqref{eq:Jd} are consistent with the integral representation~\eqref{eq:fsRint} is harder and  relegated to  Appendix~\ref{app:fsclassequiv}.

Both ways of calculating $f_{\text{s},d}^{\text{R}}$ used above can be generalized to determine the quantum corrections of $\varphi_{\text{s},d}^{\text{R}}$. To generalize the second, note that the Fourier transform with respect to $\bm{y}$ and $\tau$ of the propagator $\langle\bm{\psi}(\bm{y},z,\tau)\cdot\bm{\psi}^*(\bm{0},z',0)\rangle/2n$  associated with the action~\eqref{eq:S0} can be written as
\begin{equation}
g_\rho(\bm{p};z,z')=\frac{2m}{\hbar^2}\,\hat{G}[\kappa_{p,\rho}(\xi,\lamth);z,z';c]
\end{equation}
with
\begin{equation}\label{eq:kappaprho}
\kappa_{p,\rho}(\xi,\lamth)=\sqrt{p^2+\xi^{-2}-\rmi8\pi^2\rho/\lamth^2}.
\end{equation}
We thus arrive at the expansion
\begin{eqnarray}
\lefteqn{\varphi^{\text{R}}_{\text{s},d}(T,\mu,c)}&&\nonumber\\ &=&\varphi_{\text{s},d}(T,\mu,0)+f^{\text{R}}_{\text{s},d}(\xi,c)\nonumber\\
&&\strut+2c^{d-1}K_{d-1}\sum_{\rho=1}^\infty\Re J_d\Big[c\big(\xi^{-2}-\rmi 8\pi^2\rho/\lamth\big)^{-1/2}\Big].\nonumber\\ 
\end{eqnarray}

 Here, the first (${c=0}$) term can be expanded  about $\mu_{\text{c}}=0$ to obtain contributions analytic in $\mu_{\text{c}}-\mu$. The second term, $f_{\text{s},d}^{\text{R}}$, describes the asymptotic critical behavior. Finally, the sum $\sum_{\rho=1}^\infty\ldots$ contains the quantum corrections. They are exponentially small. 

\begin{figure}[htbp]
\begin{center}
\includegraphics[width=0.95\columnwidth]{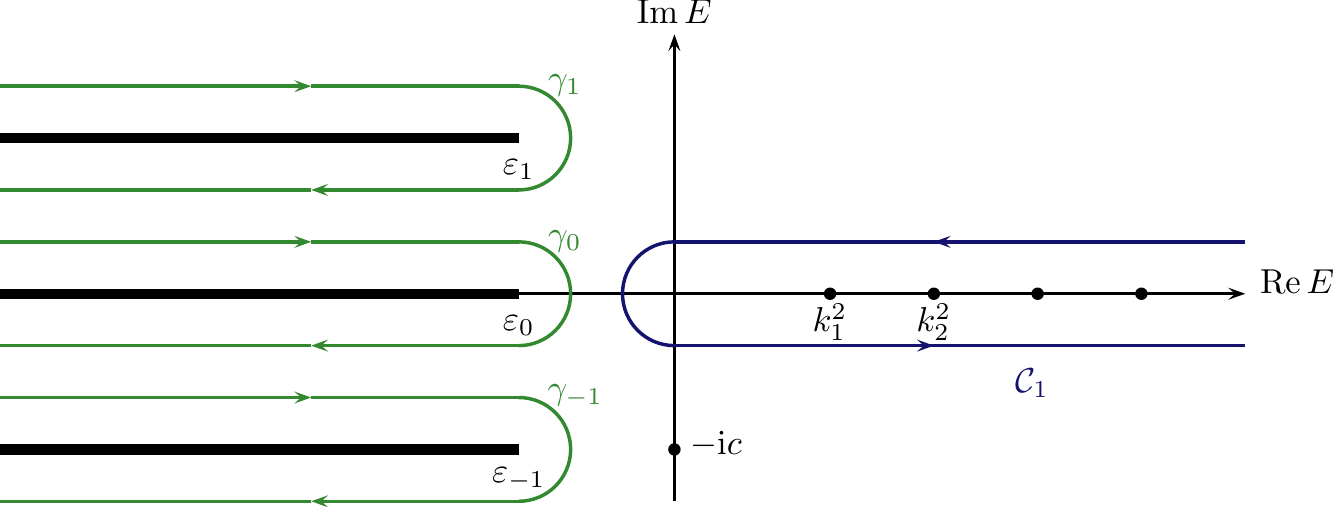}
\caption{Deformation of the contour $\mathcal{C}_1$ into the union of contours $\gamma_j,j\in \mathbb{Z}$ plus a circle around the isolated pole at $E=-\rmi c$.}
\label{fig:fig3DR16b}
\end{center}
\end{figure}

An alternative way of calculating $\varphi^{\text{R}}_{\text{s},d}(T,\mu,c)$ is to go back to the  integral representation of  $\varphi_{\text{s},d}^{\text{R}}$ in terms of $T_{\text{s}}(E;c)$:
\begin{align}
&\varphi_{\text{s},d}^{\text{R}}(T,\mu,c)\nonumber\\ &=-\lamth^{1-d}\int_{\mathcal{C}_1}\frac{\rmd{E}}{2\pi\rmi}\Li_{\frac{d+1}{2}}\Big(\rme^{-\lamth^2\frac{E+\xi^{-2}}{4\pi}}\Big)\,T_{\text{s}}(E;c).
\end{align}
Here $\mathcal{C}_1$ is the contour in the complex $E$~plane depicted in Fig.~\ref{fig:fig3DR16b}. The branch point of the polylogarithm $\Li_{(d+1)/2}(z)$ at $z=1$ yields infinitely many branch points in the complex energy plane located at 
\begin{equation}
E_j=-\xi^{-2}+\rmi 8\pi^2 \lamth^{-2}j,\;j\in \mathbb{Z}.
\end{equation} 
The thick lines in the figure denote the associated branch cuts. 

We now wish to deform the contour $\mathcal{C}_1$ into the union of contours $\mathop{\cup}\limits_{j=-\infty}^\infty\gamma_j$. In order to be able to do this, we must first ensure that the integrand of $\int_{\mathcal{C}_1}\rmd{E}$ decays sufficiently fast at $E=\infty$. This can be achieved by adding and subtracting from $T_{\text{s}}(E;c)$ its  asymptote for $E\to\infty$ given by $T_{\text{s}}(E;0)=(4E)^{-1}$. We can then deform $\mathcal{C}_1$  in the intended fashion. The polylogarithms have jump discontinuities across the branch cuts  separating the  upper and lower rims of the contours $\gamma_j$. These discontinuities  are implied by the jump
\begin{equation}\label{eq:Lijumps}
\Li_{\frac{d+1}{2}}(z)\big|_{z=x-\rmi 0}^{z=x+\rmi 0}=\frac{2\pi \rmi\ln^{(d-1)/2}(x)}{\Gamma[(d+1)/2]} \text{ for }x>1.
\end{equation}
The functions $T_s(E;c)-T_s(E,0)$ are continuous across these branch cuts since $\sqrt{E}$  (whose only branch cut is along the positive real axis) has this property. We thus obtain
\begin{eqnarray}\label{eq:varphisdqc}
\lefteqn{\varphi_{\text{s},d}^{\text{R}}(T,\mu,c)}&&\nonumber\\ &=&\frac{-1}{4}\lamth^{1-d}\,\Li_{\frac{d+1}{2}}\big(\rme^{\beta\mu}\big)- \frac{1}{\Gamma\big(\frac{d+1}{2}\big)}\sum_{j=-\infty}^\infty\int_{-\infty}^0 \rmd{u}\bigg\{\nonumber\\ &&
  \Big(\frac{-u}{4\pi}\Big)^{\frac{d-1}{2}}
\bigg[ \frac{1}{2E}\frac{c}{\rmi\sqrt{E}-c}\bigg]_{E=\ev_j+u}\bigg\},
\end{eqnarray}
where $\Im\sqrt{E}>0$. The integrals $\int_{-\infty}^0\rmd{u}u\ldots$ are convergent for $d<2$ and defined for $d>2$ by analytic continuation. 

The residual potential $\varphi_{\text{res},d}^{\text{R}}$ can be computed in a similar fashion using
\begin{eqnarray}\label{eq:varphis}
\varphi_{\text{res},d}^{\text{R}}(T,\mu,c_1,c_2)&=&-\lamth^{1-d} \int_{\mathcal{C}_1}\frac{\rmd{E}}{2\pi \rmi}\Big[\Li_{\frac{d+1}{2}}\big(\rme^{-\lamth^2\frac{E+\xi^{-2}}{4\pi}}\big)\nonumber\\ &&\times T_{\text{res}}(E;D,c_1,c_2)\Big].
\end{eqnarray}
Since $T_{\text{res}}$ decays exponentially as $E\to\infty$, the integrand needs no subtraction to deform the contour $\mathcal{C}_1$ into $\cup_{j=-\infty}^\infty\gamma_j$. Exploiting again Eq.~\eqref{eq:Lijumps}, one finds that the scaling function $\Upsilon_d^{\text{R}}$ can be written as
\begin{eqnarray}\label{eq:varphiRres}
\lefteqn{\Upsilon_d^{\text{R}}(x_\lambda,x_\xi,\scc_1,\scc_2)}&& \nonumber\\ &=&\frac{-(4\pi)^{(1-d)/2}}{\Gamma[(d+1)/2]}\sum_{j=\infty}^\infty\int_{-\infty}^0\rmd{u} \big[(-u)^{(d-1)/2}\nonumber\\ &&\times 
T_{\text{res}}(u-x_\xi^2+\rmi 8\pi^2 j x_\lambda^2);1,\scc_1,\scc_2)\big].
\end{eqnarray}

As we show in Appendix~\ref{app:UpsR}, this result can be transformed into
\begin{eqnarray}\label{eq:UpsRresd}
\lefteqn{\Upsilon_d^{\text{R}}(x_\lambda,x_\xi,\scc_1,\scc_2)}&&\nonumber\\ &=&K_{d-1} \int_0^\infty\rmd{p}\,p^{d-2}\bigg\{g_{\scc_1,\scc_2}\big[\kappa_p(1/x_\xi)\big]\nonumber\\
&&\strut + 2\sum_{\rho=1}^\infty\Re \,g_{\scc_1,\scc_2}\big[\kappa_{p,\rho}(1/x_\xi,1/x_\lambda)\big]\bigg\},
\end{eqnarray}
where
\begin{equation}\label{eq:gc1c2}
g_{\scc_1,\scc_2}(\kappa)=\ln\left[1-\frac{(\kappa-\scc_1)(\kappa-\scc_2)}{(\kappa+\scc_1)(\kappa+\scc_2)}\,\rme^{-2\kappa}\right]
\end{equation}
and $\kappa_{p,\rho}(\xi,\lamth)$ was defined in Eq.~\eqref{eq:kappaprho}. In Appendix~\ref{app:UpsR}, we also present a  somewhat easier, alternative calculation, which directly leads to Eq.~\eqref{eq:UpsRresd}.

The first contribution to $\Upsilon_d^{\text{R}}$, namely, the $\rho=0$ analog of the remaining ones,  is the classical scaling function
 \begin{equation}\label{eq:ThetaR}
\Theta^{\text{R}}_d(x_\xi,\mathsf{c}_1,\mathsf{c}_2)=K_{d-1} \int_0^\infty\rmd{p}\,p^{d-2}\,g_{\scc_1,\scc_2}[\kappa_p(1/x_\xi)].
\end{equation}
In the special case $x_\xi=0$, it reduces to its  analog at the bulk critical point, the scale-dependent amplitude 
\begin{eqnarray}\label{eq:DeltaRdef}
\Delta^{\text{R}}_d(\mathsf{c}_1,\mathsf{c}_2)&\equiv&\Theta^{\text{R}}_d(0,\mathsf{c}_1,\mathsf{c}_2)\nonumber\\
&=&K_{d-1}\int_0^\infty\rmd{p}\,p^{d-2}\,g_{\scc_1,\scc_2}(p)
\end{eqnarray}
obtained in \cite{SD08,DS11,EGJK08}. Likewise, Eqs.~\eqref{eq:ThetadDDa}--\eqref{eq:ThetadDNb} can be recovered from the result~\eqref{eq:ThetaR} in the cases $x_\xi>0$ and $(\scc_1,\scc_2)=(\infty,\infty),(0,0)$ and $(\infty,0)$.

\subsection{Quantum corrections to the critical behavior}

The results given in Eqs..~\eqref{eq:UpsRresd} and \eqref{eq:gc1c2} enable us to determine the form of  the leading quantum corrections.  They are associated with the contributions from the Matsubara frequencies $\omega_\rho$ with $\rho=\pm 1$. Let 
\begin{eqnarray}
\gamma(x_\xi,x_\lambda)&\equiv&\kappa_{0,-1}(1/x_\xi,1/x_\lambda)=\sqrt{x_\xi^2+\rmi 8\pi^2 x_\lambda^2}\nonumber\\
&=& |x_\xi^2+\rmi 8\pi^2 x_\lambda^2|^{1/2}\rme^{\rmi\arctan (8\pi^2x_\lambda^2/x_\xi^2)/2}.\qquad
\end{eqnarray}
We expand the function $g_{\scc_1,\scc_2}(\kappa_{p,-1})$ to linear order in $\rme^{-2\kappa_{p,-1}}$, write the corresponding expansion term as $A\,\rme^{-2\gamma(1+2p^2/\gamma^2)}$, and then expand the coefficient to $\Or(\gamma^{-1},p^4/\gamma^3)$. Upon performing the $p$~integral, we arrive at the expansion
\begin{eqnarray}
\lefteqn{\Upsilon_d^{\text{R}}(x_\lambda,x_\xi,\scc_1,\scc_2)-\Theta_d^{\text{R}}(x_\lambda,x_\xi,\scc_1,\scc_2)}\nonumber\\ &\mathop{=}\limits_{\gamma\to\infty}&-2\Re\bigg\{\left(\frac{\gamma}{4\pi}\right)^{\frac{d-1}{2}}\rme^{-2\gamma}\left[1+\Or\left(\rme^{-2\gamma}\right)\right]\nonumber\\&&\times\bigg[1+\frac{d^2-1}{16\gamma}-2\frac{\scc_1+\scc_2}{\gamma}+\Or\left(\gamma^{-2}\right)\bigg] \bigg\}.\quad
\end{eqnarray}
At the bulk critical point, where $x_\xi=0$, we have $|\gamma|=2^{3/2}\pi x_\lambda$. Thus, the quantum corrections to the Casimir amplitude $\Delta_d^{\text{R}}(\scc_1,\scc_2)$ are down by a factor $\exp(-2^{5/3}\pi D/\lamth)$.

An analogous analysis of quantum corrections can be made in the cases of the PBCs and ABCs. To see this, note that the classical scaling functions $\Theta_d^{\text{P,A}}$ can be written in a form analogous to Eqs.~\eqref{eq:ThetaR} using integration by parts. Adding the quantum corrections resulting from the Matsubara frequencies $\omega_\rho\ne0$ then gives
\begin{eqnarray}\label{eq:upsperresd}
\lefteqn{\Upsilon_d^{\text{P/A}}(x_\lambda,x_\xi)/n}&&\nonumber\\ &=&\pm 2K_{d-1} \int_0^\infty\rmd{p}\,p^{d-2}\bigg\{\ln\big[1\mp\rme^{-\kappa_p(1/x_\xi)}\big]\nonumber\\
&&\strut + 2\sum_{\rho=1}^\infty\Re \ln\big[1\mp\rme^{-\kappa_{p,\rho}(1/x_\xi,1/x_\lambda)}\big]\bigg\},
\end{eqnarray}
where the upper (lower) signs refer to $\text{BC}=\text{P}$ and $\text{A}$, respectively. It follows that the leading quantum corrections are of the form
\begin{eqnarray}
\lefteqn{\Upsilon_d^{\text{P,A}}(x_\lambda,x_\xi)/n-\Theta_d^{\text{P,A}}(x_\lambda,x_\xi)/n}\nonumber\\ &\mathop{=}\limits_{\gamma\to\infty}& -4\,\Re\bigg\{\left(\frac{\gamma}{2\pi}\right)^{\frac{d-1}{2}}\rme^{-\gamma}\left[1+\frac{d^2-1}{8\gamma}+O\left(\gamma^{-2}\right)\right]\nonumber\\&&\times \left[1+O\left(\rme^{-\gamma}\right)\right]\bigg\}.
\end{eqnarray}
Thus, the quantum corrections to the amplitudes $\Delta_d^{\text{P/A}}$ are smaller by a factor $\exp(-2^{3/2}\pi D/\lamth)$, the decay length $\lamth/2^{3/2}$ is twice as large as in the cases of DDBCs, NNBCs, DNBCs, and RBCs.

\section{Interacting $n$-component Bose gas \label{sec:intBG}}

\subsection{The limit $n\to\infty$}

We now turn to the interacting $n$-component  Bose gas. In order to investigate its  limit ${n\to\infty}$, we use the Hubbard-Stratonovich transformation 
\begin{equation}\label{eq:HStransf}
\rme^{-(\ub/2n)|\psi|^4}=\sqrt{\frac{n}{2\pi \ub}}\int_{-\infty}^\infty\rmd\chi\,
\rme^{-\rmi \,\chi|\,\psi|^2-(n/2\ub)\chi^2}\quad
\end{equation}
to rewrite the interaction term of the action~\eqref{eq:S}. It follows that $\rme^{-S[\bm{\psi}*,\bm{\psi}]/\hbar}$ can be written as a  functional integral $\int\mathcal{D}[\chi]\,\exp\big(-S_{\text{eff}}[\bm{\psi}^*,\bm{\psi},\chi]/\hbar\big)$ over the real-valued field $\chi(\bm{x},\tau)$. The effective action is given by 
\begin{eqnarray}\label{eq:Seff1}
S_{\text{eff}}[\bm{\psi}^*,\bm{\psi},\chi]&=&\int_{\tau,\mathfrak{V}}\bigg[\psi^*_\alpha\bigg(\hbar\partial_\tau-\mu+\rmi\chi-\frac{\hbar^2}{2m}\nabla^2\bigg)\psi_\alpha
 \nonumber\\ &&\strut
 +\frac{n}{2\ub }\chi^2 \bigg],
\end{eqnarray}
up to an unimportant constant. Here, $\int_{\tau,\mathfrak{V}}$ stands for $\int_0^{\beta\hbar}\rmd{\tau}\int_{\mathfrak{V}}\rmd^dx$. Upon integrating out the fields $\bm{\psi}^*$ and $\bm{\psi}$ and introducing the potential 
\begin{equation}\label{eq:Vpotdef}
\mathcal{V}(\bm{x},\tau)=\rmi\chi(\bm{x},\tau)-\mu,
\end{equation}
 one arrives at
\begin{eqnarray}\label{eq:Seff}
\lefteqn{\ln\int\mathcal{D}[\bm{\psi}^*,\bm{\psi},\chi]\,\rme^{-S_{\text{eff}}[\bm{\psi}^*,\bm{\psi}]/\hbar}+\text{const}}&&\nonumber\\
&=&-n\int_{\tau,\mathfrak{V}}\Big\{\Big\langle\bm{x},\tau\Big|\ln\Big[\hbar\partial_\tau+\mathcal{V}-\frac{\hbar^2}{2m}\nabla^2\Big]\Big|\bm{x},\tau\Big\rangle
\nonumber\\&&
-\frac{ 1}{2 \ub}(\mu+\mathcal{V})^2\Big\}/\hbar,
\end{eqnarray}
where  a self-explanatory Dirac notation is used.

The explicit factor of $n$ in the result tells us that the functional integral over $\chi$  for $n\to\infty$ can be calculated by evaluating the integrand at the stationary point. The stationary potential $\mathcal{V}_*(\bm{x})=\mathcal{V}(\bm{x})$ (where the subscript asterisk indicates values at the stationary point) is independent of $\tau$ by translation invariance along the time direction. Since we also have spatial translation invariance along the $\bm{y}$~directions, it is sufficient to consider potentials that depend  on $z$  or  are independent of $z$ in the cases of free BCs and PBCs, respectively. For such potentials
\begin{equation}\label{eq:Vv}
\mathcal{V}(\bm{y},z)=\mathcal{V}(z)\equiv \frac{\hbar^2}{2m}v(z)
\end{equation}
the eigenvalues of the operator  $v(z)-\nabla^2$ are of the form $p^2+\ev_\nu$ where $\bm{p}$ denotes the wave vector conjugate to $\bm{y}$ and $\ev_\nu\equiv \ev_\nu^{\text{BC}}$, $\nu=1,2,\dotsc,\infty$, are the eigenvalues of the operator $-\partial_z^2+v(z)$ on the interval $[0,D]$ subject to the BCs considered.

 Let us introduce the grand partition function
\begin{equation}
\Xi_{d,n}^{\text{BC}}(T,\mu,D,L,\ub)=\int\mathcal{D}[\bm{\psi}^*,\bm{\psi}]\int\mathcal{D}[\chi]\,\rme^{-S_{\text{eff}}[\bm{\psi}^*,\bm{\psi},\chi]/\hbar}
\end{equation}
 by analogy with Eq.~\eqref{eq:Xi}. In the large-$n$ limit the associated grand potential per hyperarea $A$ and number $n$ of components
 \begin{equation}
\varphi_{\infty,d}^{\text{BC}}(T,\mu,\ub,D,L)\equiv=-\frac{1}{A}\lim_{n\to\infty}\frac{1}{n}\ln\Xi_d^{\text{BC}}(T,\mu,\ub,D,L)
\end{equation}
is given by the maximum of the functional 
\begin{align}
&\Phi([\mathcal{V}];T\mu,\ub,D,L)=\psum\int_0^D\rmd{z}\bigg\{-\frac{\beta}{2\ub}[\mathcal{V}(z)+\mu]^2\nonumber\\ &\strut +\int_0^{\beta\hbar}\frac{\rmd{\tau}}{\hbar}\Big\langle z,\tau\Big|\ln\Big[\hbar\partial_\tau+\mathcal{V}(z)+\frac{\hbar^2}{2m}(p^2-\partial_z^2)\Big]\Big| z,\tau \Big\rangle\bigg\},
\end{align}
where we have chosen PBCs along the $\bm{y}$~directions and $\bm{p}$ is a (${d-1}$)-dimensional wave vector with $\frac{L}{2\pi}\,\bm{p}\in\mathbb{Z}^{d-1}$. Functional differentiation with respect to $\mathcal{V}(z)$ yields the stationarity condition
\begin{equation}\label{eq:sce}
\ub\psum g_{\bm{p}}(z,z,0-)=\mathcal{V}_*(\bm{x})+\mu,
\end{equation}
where $g_{\bm{p}}(z,z',\tau-\tau')$ is a Matsubara Green's function satisfying
\begin{equation}\label{eq:Mg}
{\left[\hbar\partial_\tau+\mathcal{V}_*+\frac{\hbar^2}{2m}\big(p^2-\partial_z^2\big)\right]}g_{\bm{p}}(z,z',\tau)=\delta(\tau)\,\delta(z-z').
\end{equation}
Let $\ev_\nu$ and $\ef_\nu(z)$  be the eigenvalues and orthonormalized eigenfunctions  of the Schr\"odinger equation
\begin{equation}\label{eq:Seq}
\left[-\partial_z^2+v_*(z)-\ev_\nu\right]\ef_\nu(z)=0,
\end{equation}
and $n_{\bm{p},\nu}$ denote the occupation number
\begin{equation}
n_{\bm{p},\nu}=\left\{\exp\left[\frac{\beta\hbar^2}{2m} (p^2+\ev_\nu)\right]-1\right\}^{-1}.
\end{equation}
Then we have
\begin{eqnarray}
g_{\bm{p}}(z,z',\tau) &=&\sum_\nu \Big\{\ef_\nu (z)\ef^*_\nu (z')\,\big[n_{\bm{p},\nu}\Htheta(-\tau)\nonumber\\ &&\strut+(1+n_{\bm{p},\nu})\Htheta(\tau)\big]\,\rme^{-\frac{\hbar}{2m}(\ev_\nu+p^2)\tau}\Big\},\qquad
\end{eqnarray}
and the self-consistency Eq.~\eqref{eq:sce} becomes
\begin{equation}\label{eq:sce2}
\mathcal{V}_*(z)+\mu
=\ub\sum_\nu|\ef_\nu (z)|^2\psum\,n_{\bm{p},\nu}.
\end{equation}
Furthermore, the contribution to the functional $\Phi_d([\mathcal{V}];T,\mu,D,L,\ub)$ associated with the integral $\int\rmd{\tau}$ is nothing else than the reduced grand potential of a set of noninteracting bosons with single-particle energies $\frac{\hbar^2}{2m}(p^2+\ev_\nu+\mu)$. Accordingly, we have
 \begin{eqnarray}
\Phi_d([\mathcal{V}];T,\mu,\ub,D,L)&=&
\psum\sum_\nu\ln\big[1-\rme^{-\frac{\beta\hbar^2}{2m}(p^2+\ev_\nu)}\big]\nonumber\\
&&\strut-\frac{\beta}{2\ub}\int_0^D\rmd{z}\,[\mathcal{V}(z)+\mu]^2.
\end{eqnarray}

We now wish to take the limit $A=L^{d-1}\to\infty$. To allow for macroscopic occupancy of $\bm{p}=\bm{0}$ states, we separate the ${\bm{p}=\bm{0}}$ from $\sum_{\bm{p}}$ using the asymptotic equivalence
\begin{equation}
\psum f(\bm{p})\mathop{\aseq}_{A\to\infty}A^{-1}f(\bm{0})+\pint f(\bm{p}).
\end{equation}
The self-consistency condition of Eqs.~\eqref{eq:sce} and \eqref{eq:sce2} thus becomes
\begin{eqnarray}\label{eq:sce3}
\mathcal{V}_*(z)+\mu&\mathop{=}\limits_{A\to\infty}&\ub\sum_\nu|\ef_\nu(z)|^2\,\lamth^{1-d}\,\Li_{\frac{d-1}{2}}\big(\rme^{-\frac{\beta\hbar^2}{2m}\ev_\nu}\big)\nonumber\\&&\strut +\frac{\ub}{A}\sum_\nu |\ef_\nu(z)|^2\,\frac{1}{\rme^{\frac{\beta\hbar^2}{2m}\ev_\nu}-1}
\end{eqnarray}
and for the ${A\to\infty}$ limit of the grand potential we find
\begin{eqnarray}\label{varphiinfD}
\lefteqn{\varphi^{\text{BC}}_{\infty,d}(T,\mu,\ub,D)\equiv \varphi^{\text{BC}}_{\infty,d}(T,\mu,\ub,D,\infty)}\nonumber \\ &=&-\lamth^{1-d}\sum_\nu\Li_{\frac{d+1}{2}}\Big(\rme^{-\frac{1}{4\pi}\lamth^2\ev_\nu}\Big)
 -\frac{\beta}{2\ub}\int_0^D\rmd{z}\big[\mu+\mathcal{V}_*(z)\big]^2\nonumber\\ &&\strut +\lim_{A\to\infty}\frac{1}{A}\sum_\nu\ln\Big(1-\rme^{-\frac{1}{4\pi}\lamth^2\ev_\nu}\Big).
\end{eqnarray}
The contributions from the ${\bm{p}=\bm{0}}$ terms to both equations [last terms in Eqs.~\eqref{eq:sce3} and \eqref{varphiinfD}] vanish in the  limit $A\to\infty$.

It is an easy matter to check that these equations reduce to those of \cite{NJN13} if we choose PBCs. To see this, note that in this case the self-consistent potential is independent of $z$ by translational invariance, we have 
\begin{eqnarray}
\mathcal{V}_*^{\text{P}}&\equiv& \mathcal{V}_*^{\text{P}}(T,\mu,\ub,D)=\frac{\hbar^2}{2m}v_*^{\text{P}}(T,\mu,\ub,D),\nonumber\\ 
\ev^{\text{P}}_\nu&=&v_*^{\text{P}}+(k^{\text{P}}_\nu)^2,
\end{eqnarray}
where $k^{\text{P}}_\nu$ are the discrete $k$~values given in  Eq.~\eqref{eq:per}. We insert these results into Eqs.~\eqref{eq:sce3} and \eqref{varphiinfD} along with the eigenfunctions $\ef^{\text{P}}_\nu(z)$ given in Eq.~\eqref{eq:per}. Taking into account that the variables $s_0$ and $a$ of \cite{NJN13} correspond to $-\beta \mathcal{V}^{\text{P}}$ and $\ub$, respectively, one recovers the corresponding equations of \cite{NJN13}.
\subsection{Bulk properties}\label{sec:bulkprop}
The foregoing statements carry over to the bulk quantities
\begin{equation}
\varphi_{\infty,\text{b},d}(T,\mu,\ub)\equiv \lim_{D\to\infty}D^{-1}\,\varphi^{\text{BC}}_{\infty,d}(T,\mu,\ub,D)
\end{equation}
and
\begin{eqnarray}
\mathcal{V}_{\text{b},*}&\equiv&\mathcal{V}_{\text{b},*}(T,\mu,\ub)=\mathcal{V}_*^{\text{P}}(T,\mu,{D=\infty},\ub)\nonumber\\ &=&\mathcal{V}_*^{\text{BC}}(z=\infty;T,\mu,{D=\infty},\ub),\nonumber\\ &&\text{BC}=\text{DD},\text{NN},\text{DN},\text{R}.\qquad
\end{eqnarray}
Note that $\mathcal{V}_{\text{b},*}$ can be defined either as the $D\to\infty$ limit of $\mathcal{V}^{\text{P}}_*$ or in terms of the $z\to\infty$ limit of the potential $\mathcal{V}_{*}|_{D=\infty}$ for the semi-infinite case with free boundary conditions.

The equations these quantities satisfy, 
\begin{eqnarray}\label{eq:varphiinfb}
\varphi_{\infty,\text{b},d}(T,\mu,\ub)&=&-\lamth^{-d}\,\Li_{\frac{d+2}{2}}\Big(\rme^{-\beta\mathcal{V}_{\text{b},*}}\Big) -\frac{\beta}{2\ub}\big(\mu+\mathcal{V}_{\text{b},*}\big)^2\nonumber\\ &&\strut +\lim_{V\to\infty}\frac{1}{V}\ln\Big(1-\rme^{-\beta\mathcal{V}_{\text{b},*}}\Big)\quad
\end{eqnarray}
and
\begin{equation}\label{eq:bulksce}
\mathcal{V}_{\text{b},*}+\mu\mathop{\aseq}_{V\to\infty}\ub\,\lamth^{-d}\,\Li_{d/2}\Big(\rme^{-\beta\mathcal{V}_{\text{b},*}}\Big)+\frac{\ub}{V}\frac{1}{\rme^{\beta\mathcal{V}_{\text{b},*}}-1},
\end{equation}
are again in accordance with those of \cite{NJN13}. The easiest way to obtain these bulk equations is to choose PBCs. Alternatively, one can consider the semi-infinite case and investigate the limit $z\to\infty$.

For later use, let us also mention that the following results (obtained in \cite{NJN13}) follow for $2<d<4$ in a straightforward fashion  from the above bulk equations: 
\begin{enumerate}
\item[(i)] The critical line $\mu_{\text{c}}(T,\ub)$ across which the bulk transition occurs is given by
\begin{equation}\label{eq:muc}
\mu_{\text{c}}(T,\ub)=\lamth^{-d}\ub\,\zeta(d/2).
\end{equation}
\item[(ii)] The bulk potential $\mathcal{V}_{\text{b},*}$ vanishes on the critical line $\mu_{\text{c}}(T,\ub)$ and in the bulk ordered phase in the thermodynamic limit $V\to\infty$.
\item[(iii)] The bulk grand potentials $\varphi^<_{\text{b},d}(T,\mu)$ and $\varphi^>_{\text{b},d}(T,\mu)$ in the bulk disordered and bulk ordered phases can be written as
\begin{eqnarray}\label{eq:inftybd<}
\varphi^<_{\infty,\text{b},d}(T,\mu,\ub)&=&-\big[\lamth(T)\big]^{-d}\,\Li_{d/2+1}\big(\rme^{-\beta\mathcal{V}_{\text{b},*}}\big)\nonumber\\ &&\strut-\frac{\beta}{2\ub}(\mathcal{V}_{\text{b},*}+\mu)^2
\end{eqnarray}
and
\begin{equation}\label{eq:inftybd>}
\varphi^<_{\infty,\text{b},d}(T,\mu,\ub)=-\frac{\beta}{2\ub}\,\mu^2-\big[\lamth(T)\big]^{-d}\,\zeta(d/2+1),
\end{equation}
respectively.
\item[(iv)] Analysis of the bulk equations~\eqref{eq:bulksce} and \eqref{eq:varphiinfb} yields the limiting behaviors 
\begin{equation}
\beta\mathcal{V}_{\text{b},*}[T,\mu_{\text{c}}(1+\delta\mu),\ub]\mathop{=}_{\mu\to \mu_{\text{c}}-}\left[\frac{\zeta(d/2)}{-\Gamma(1-d/2)}\,(-\delta\mu)\right]^{\gamma_\infty}
\end{equation}
and
\begin{align}
&\varphi_{\infty,\text{b},d}[T,\mu_{\text{c}}(1+\delta\mu),\ub]-\varphi_{\infty,\text{b},d}(T,\mu_{\text{c}},\ub)\nonumber\\ &\mathop{\aseq}\limits_{\delta\mu\to 0-}-\frac{\Gamma(-d/2)}{\lamth^d(T)}\bigg[\frac{\zeta(d/2)}{-\Gamma(1-d/2)}(-\delta\mu)\bigg]^{2-\alpha_\infty}+\Or(\delta\mu)\quad
\end{align}
for $\delta\mu\equiv (\mu/\mu_{\text{c}}-1)\to 0^-$, where $\gamma_\infty=2\nu_\infty=2/(d-2) \equiv\gamma_{\text{id}}$ and $\alpha_\infty=(4-d)/(d-2)=\alpha_{\text{id}}$ are the familiar  $(n=\infty)$ critical exponents of the spherical model. 
\end{enumerate}

The linear scaling field $\delta\mu$ varies linearly in $\delta T$. However, unlike its Gaussian analog $\alpha_{\text{G}}$, the exponent $\alpha_\infty$ is negative so that $\delta\mu \asprop \delta T$ as $\delta T\to 0$. Therefore, the constraint of constant $\rho$ does not lead to a Fisher renormalization of the $(n=\infty)$ critical exponents. 

Comparison of the $n\to\infty$ action with that of the  ideal Bose gas shows that the bulk correlation length $\xi$ is given by the analog of Eq.~\eqref{eq:xiid} obtained by the replacement $-\mu\to V_{\text{b},*}$, i.e.,
\begin{equation}\label{eq:ninftybc}
\xi=\frac{\hbar}{\sqrt{2m\mathcal{V}_{\text{b},*}}}=\frac{\lamth}{2\sqrt{\pi}}\big(\beta \mathcal{V}_{\text{b},*}\big)^{-1/2}.
\end{equation}

As an immediate consequence, one obtains for the static pair correlation function in the disordered phase
\begin{eqnarray}\label{eq:IdBGcf}
g_{\text{b},d}(\bm{x};\lamth,\xi)&=&\lim_{n\to\infty}\frac{1}{n}\big\langle\psi^*_\alpha(\bm{x},\tau)\psi_\alpha(\bm{0},\tau-)\big\rangle_{\text{b},d}\nonumber  \\ &=&\int_{\bm{q}}^{(d)}\frac{\rme^{\rmi\bm{q}\cdot\bm{x}}}{\rme^{\beta(\mathcal{V}_{\text{b},*}+\hbar^2q^2/2m)}-1}\nonumber \\ &=&\lamth^{-d}\sum_{s=1}^\infty s^{-d/2}\rme^{-s\lamth^2/4\pi\xi^2-\pi x^2/s\lamth^2}.\nonumber\\
\end{eqnarray}
To determine its  asymptotic behavior in the regime $\lamth\ll x,\xi$, one can rescale according to Eq.~\eqref{eq:clresc} and  take the limit
\begin{eqnarray}\label{eq:gbgbcl}
\lefteqn{\lim_{\lamth\to 0}\frac{\lamth^2}{2\pi}\,g_{\text{b},d}(\bm{x};\lamth,\xi)}&&\nonumber\\&=&2(2\pi)^{-d/2}\,(x\xi)^{-\frac{d-2}{2}}\,K_{\frac{d-2}{2}}(x/\xi),\nonumber\\ &\equiv&2g^{\text{cl}}_{\text{b},d}(\bm{x};\xi),
\end{eqnarray}
which eliminates quantum corrections of order $\Or\big(\rme^{-\text{const}\,x/\lamth}\big)$. The right-hand side is twice the propagator $g^{\text{cl}}_{\text{b},d}(\bm{x};\xi)\equiv \lim_{n\to\infty}\sum_{\alpha_2=1}^{2n}\langle\bm{\phi}_{\alpha_2}(\bm{x})\phi_{\alpha_2}(\bm{0})\rangle_{\text{b},d}/2n$ of the  classical $O(2n)$ $\phi^4$ theory in the disordered phase, as it should. This shows that $\xi$ is the true correlation length, namely, the scale on which this function decays exponentially in the large-distance limit  $x\to\infty$.

All  above-mentioned results for the bulk critical behavior of the imperfect Bose gas and our interacting $n$-component Bose gas are in accordance with the fact that this behavior is representative of the universality class of the $O(2n)$ $\phi^4$ model in the limit $n\to\infty$.  For PBCs, this identification of the universality class for the scaling behavior of both of these  Bose~gas models near bulk criticality carries over to the  case of finite thickness $D$. In the next subsection we explicitly verify that the asymptotic scaling behaviors of the residual grand potential $\varphi_{\infty,d}^{\text{P}}$ and the associated Casimir force $\mathcal{F}^{\text{P}}_{C,\infty}$ near the bulk transition are indeed described by the scaling functions of the classical $O(2n)$ model in the $n\to\infty$ limit.

\subsection{Scaling functions for periodic boundary conditions}

We begin by considering the case $\mu<\mu_{\text{c}}$ where both the bulk system and the $D<\infty$ strip are disordered whenever $T>0$. Let us generalize  Eqs.~\eqref{eq:fresUp}, \eqref{eq:betaFC}, \eqref{eq:UpsThetarel},  and \eqref{eq:varhetaBCdef} to the present interacting case. The interaction constant $\ub$ or, equivalently, the rescaled interaction constant $g$ defined in Eq.~\eqref{eq:gdef}, gives rise to a further dimensionless variable. As dimensionless variables, we choose
\begin{equation}
x_\lambda=D/\lamth,\quad x_\xi=D/\xi,\quad x_g=gD^{4-d},
\end{equation}
where $\xi$ now denotes the bulk correlation length~\eqref{eq:ninftybc} rather than its ideal Bose~gas counterpart $\xi_{\text{id}}$. We can then write the ${n=\infty}$~analogs of  Eqs.~\eqref{eq:fresUp} and \eqref{eq:betaFC} as
\begin{equation}\label{eq:fresUpninfty}
\varphi_{\infty,\text{res},d}^{\text{BC}}(T,\mu,\ub,D)=D^{-(d-1)}\,\Upsilon_{\infty,d}^{\text{BC}}(x_\lambda,x_\xi,x_g).
\end{equation}
and
\begin{equation}\label{eq:betaFCninfty}
\beta\mathcal{F}^{\text{BC}}_{\infty,C}(T,\mu,\ub,D)=D^{-d}\,\mathcal{Y}_{\infty,d}^{\text{BC}}(x_\lambda,x_\xi,x_g),
\end{equation}
respectively. In the scaling limit of the Bose-Einstein transition, where $D$ and $\xi$ both become large compared to all other lengths (namely, $\lamth$ and $g^{-1/(4-d)}$), the behavior should simplify to
\begin{equation}\label{eq:UpsThetarelinfty}
\varphi_{\infty,\text{res},d}^{\text{BC}}(T,\mu,\ub,D)\mathop{\aseq}\limits_{D,\xi\to\infty}D^{-(d-1)}\Theta_{\infty,d}^{\text{BC}}(D/\xi)
\end{equation}
and
\begin{equation}\label{FC:varthetainfty}
\beta\mathcal{F}^{\text{BC}}_{\infty,C}(T,\mu,\ub,D)\mathop{\aseq}\limits_{D,\xi\to \infty}D^{-d}\vartheta_{\infty,d}^{\text{BC}}(D/\xi),
\end{equation}
where $\Theta_{\infty,d}$ and $\vartheta_{\infty,d}$ are scaling functions of the classical $O(2\infty)$ $\phi^4$ model. The dependence on both $\lamth$ and $g$ (or $\ub$) should drop out except from nonuniversal amplitudes such as that of $\xi$.

To determine  the residual grand potential $\varphi_{\infty,\text{res},d}^{\text{P}}$, we must compute the grand potentials $\varphi_{\infty,\text{b},d}^{\text{P}}$ and $\varphi_{\infty,d}^{\text{P}}$. The latter functions can be conveniently written in terms of the ideal Bose~gas bulk correlation function~\eqref{eq:IdBGcf} and  its finite-$D$ analog under PBCs,
\begin{eqnarray}\label{eq:varphiDgD}
\lefteqn{g^{\text{P}}_{D,d}(y,z;\lamth,\xi_D)}\nonumber\\&=&\frac{1}{D}\sum_{k=k^{\text{P}}}\pint\frac{\rme^{\rmi(\bm{p},k)\cdot(\bm{y},z)}}{\rme^{\beta[\mathcal{V}_{*}^{\text{P}}+\hbar^2(p^2+k^2)/2m]}-1} \nonumber\\ &=&\sum_{j=-\infty}^\infty g_{\text{b},d}\big[\sqrt{y^2+(z+jD)^2};\lamth,\xi_D\big].
\end{eqnarray}
Here, the summation $\sum_{k^{\text{P}}}$ is over all $k_\nu^{\text{P}}=2\pi\nu/D$ with $\nu\in \mathbb{Z}$, and 
\begin{equation}\label{eq:ninftyDc}
\xi_D=\frac{\hbar}{\sqrt{2m\mathcal{V}^{\text{P}}_{*}}}=\frac{\lamth}{2\sqrt{\pi}}\big(\beta \mathcal{V}^{\text{P}}_{*}\big)^{-1/2}
\end{equation}
means the analog of the bulk correlation length~\eqref{eq:ninftybc}.
The last line of Eq.~\eqref{eq:varphiDgD}  can be obtained via the method of images or Poisson's summation formula
\begin{equation}\label{eq:Psumform}
\frac{1}{D}\sum_{k=k^{\text{P}}} f(k)=\sum_{j=-\infty}^\infty\int_{-\infty}^\infty\frac{\rmd{k}}{2\pi}\,f(k)\,\rme^{\rmi j D k}
\end{equation}
for functions $f(k)$.

Expressed in terms of the pair correlation function $g^{\text{P}}_{d,D}$,  the grand potential $\varphi^{\text{P}}_{\infty,d}$ becomes
\begin{eqnarray}\label{eq:varphiPDf}
\varphi^{\text{P}}_{\infty,d}(T,\mu,\ub,D)/D&=&-\lamth^2\, g^{\text{P}}_{D,d+2}(0;\lamth,\xi_D)\nonumber\\ &&\strut  -\frac{\beta }{2\ub}\big(\mu+\mathcal{V}^{\text{P}}_{*}\big)^2.
\end{eqnarray}
The corresponding formula for $\varphi_{\infty,\text{b},d}(T,\mu,\ub)$, which follows upon taking the limit $D\to\infty$, should be obvious. The respective  form of the self-consistency condition follows in a straightforward fashion by equating the derivative $\partial/\partial\mathcal{V}^{\text{P}}_*$ to zero,  using its stationarity at $\mathcal{V}=\mathcal{V}^{\text{P}}_*$.   This yields
\begin{equation}\label{eq:scePD}
\mathcal{V}^{\text{P}}_*+\mu=\ub\sum_{j=-\infty}^\infty g_{b,d}( Dj;\lamth,\xi_D)
\end{equation}
and the bulk analog
\begin{equation}\label{eq:bsce}
\mathcal{V}_{\text{b},*}+\mu=\ub\, g_{b,d}( 0;\lamth,\xi).
\end{equation}
Eliminating $\mu$ from these two equations  and expressing $\ub$ in terms of the coupling constant $g$ defined in Eq.~\eqref{eq:gdef} yields an equation whose solution gives us $D/\xi_D$ as a function of $x_\xi$, $x_\lambda$, and $x_g$:
\begin{equation}
D/\xi_D=\mathcal{X}_d(x_\lambda,x_\xi,x_g),
\end{equation}
where $\mathcal{X}_d(x_\lambda,x_\xi,x_g)$ satisfies
\begin{eqnarray}\label{eq:eqforXd}
\frac{6}{x_g}\big[\mathcal{X}_d^2-x_\xi^2\big]&=&\frac{1}{2\pi x_\lambda^2}\bigg[g_{b,d}\big( 0;x_\lambda^{-1},1/\mathcal{X}_d\big)
\nonumber\\ &&\strut 
-g_{b,d}\big(0;x_\lambda^{-1},x_\xi^{-1}\big) 
\nonumber\\ &&\strut 
+2\sum_{j=1}^\infty g_{b,d}( j;x_\lambda^{-1},1/\mathcal{X}_d)\bigg].
\end{eqnarray}

In the limit $x_\lambda\to\infty$, the solution $\mathcal{X}_d$ agrees with the classical value $\mathcal{X}^{\text{cl}}_d(x_\xi,x_g)\equiv\mathcal{X}_d(0,x_\xi,x_g)$ up to exponentially small quantum corrections:
\begin{equation}
\mathcal{X}_d(x_\lambda,x_\xi,x_g)\mathop{=}_{x_\lambda\to\infty}\mathcal{X}^{\text{cl}}_d(x_\xi,x_g)+\Or\big(\rme^{-1/x_\lambda}\big),
\end{equation}
where $\mathcal{X}^{\text{cl}}_d(x_\xi,x_g)$ solves the classical analog of Eq.~\eqref{eq:eqforXd} one obtains upon taking the limit $x_\lambda\to0$. The limit value of the right-hand side is easily determined with the aid of Eq.~\eqref{eq:gbgbcl} or by replacing the sums $\sum_{s=1}^\infty$ the functions $g_{\text{b},d}$ involve by integrals $\int_0^\infty\rmd{s}$. Introducing the coefficient
\begin{equation}
A_d=-(4\pi)^{-d/2}\Gamma(1-d/2)
\end{equation}
and using the result~\eqref{eq:Thetadpera} for the ideal Bose gas scaling function $\Theta_d^{\text{P}}$,
we see that the resulting equation for  $\mathcal{X}^{\text{cl}}_d(x_\xi,x_g)$ can be written as
\begin{eqnarray}\label{eq:eqforXdcl}
\frac{6}{x_g}\big[\big(\mathcal{X}^{\text{cl}}_d\big)^2-x_\xi^2\big]&=&-2A_d\big[\big(\mathcal{X}^{\text{cl}}_d\big)^{d-2}-x_\xi^{d-2}\big]
\nonumber\\ &&\strut 
+\frac{1}{2\pi}\,\Theta^{\text{P}}_{d-2}\big(\mathcal{X}^{\text{cl}}_d\big).
\end{eqnarray}

The $x_g$ depending terms on the left-hand side of Eq.~\eqref{eq:eqforXdcl} yield corrections to scaling to $\xi_D/D$  of the form $g^{-1}D^{-(4-d)}$ with the familiar ${n=\infty}$ exponent $\omega_g=4-d$. As is discussed in some detail in \cite{DGHHRS12} and \cite{DGHHRS14}, they can be eliminated by taking the limit $g\to\infty$ \cite{fn5}. In fact,  it follows from Eq.~\eqref{eq:eqforXdcl} that the function $\mathcal{X}^{\text{cl}}_d(x_\xi,x_g)$ behaves asymptotically as
\begin{equation}
\mathcal{X}^{\text{cl}}_d(x_\xi,x_g)\mathop{=}_{x_g\to\infty}\mathcal{X}^{\text{cl}}_{d,\text{as}}(x_\xi)+\Or(1/x_g),
\end{equation}
where $\mathcal{X}^{\text{cl}}_{d,\text{as}}$ is the zero of the right-hand side of this equation. 

At ${d=3}$, where $\Theta^{\text{P}}_{d-2}$ simplifies to $-2\ln(1-\rme^{-1/x_\xi})$, this zero can be determined in closed analytic form. One finds \cite{DDG06,fn6}
\begin{eqnarray}\label{eq:x3clas}
\mathcal{X}^{\text{cl}}_{3,\text{as}}(x_\xi)&=&2\,\text{arsinh}(\rme^{x_\xi/2}/2)\nonumber\\ &=&2\ln\big[\big(\rme^{x_\xi/2}+\sqrt{4+\rme^{x_\xi}}\big)/2\big],
\end{eqnarray}
in agreement with \cite{Dan96}, \cite{Dan98}, and \cite{DDG06}.

The residual grand potential $\varphi_{\infty,\text{res},d}^{\text{P}}$ can be computed along similar lines. Upon eliminating $\mu$ in favor of $\lamth$ and $\xi$ by means of Eq.~\eqref{eq:bsce}, we can express $\varphi_{\infty,\text{res},d}^{\text{P}}$   in terms of the variables $x_\lambda$, $x_\xi$ and $x_g$ to determine the function $\Upsilon_{\infty,d}^{\text{P}}(x_\lambda,x_\xi,x_g)$. We obtain
\begin{eqnarray}
\lefteqn{\Upsilon_{\infty,d}^{\text{P}}(x_\lambda,x_\xi,x_g)}&&\nonumber\\ &=&x_\lambda^d\Big[\Li_{1+d/2}\Big(\rme^{-x_\xi^2/4\pi x_\lambda^2}\Big)-\Li_{1+d/2}\Big(\rme^{-\mathcal{X}_d^2/4\pi x_\lambda^2 }\Big)\Big]\nonumber\\ 
&&\strut+\Upsilon_d^{\text{P}}(x_\lambda,\mathcal{X}_d)-\frac{3}{2x_g}(\mathcal{X}_d^2-x_\xi^2)^2\nonumber\\&&\strut -\frac{1}{4\pi}x_\lambda^{d-2}(\mathcal{X}_d^2-x_\xi^2)\,\Li_{d/2}\big(\rme^{-x_\xi^2/4\pi x_\lambda^2}\big).
\end{eqnarray}
In the limit $x_\lambda\to\infty$, this becomes
\begin{eqnarray}
\lefteqn{\Upsilon_{\infty,d}^{\text{P}}(\infty,x_\xi,x_g)}&&\nonumber\\ &=&4\pi A_{d+2}\big[\big(\mathcal{X}^{\text{cl}}_d\big)^d-x_\xi^d\big]+\Theta_d^{\text{P}}(\mathcal{X}^{\text{cl}}_d)\nonumber\\ &&\strut +A_d x_\xi^{d-2}\big[(\mathcal{X}^{\text{cl}}_d)^2-x_\xi^2\big]-\frac{3}{2x_g}\big[(\mathcal{X}^{\text{cl}}_d)^2-x_\xi^2\big]^2.\qquad
\end{eqnarray}

Ignoring the corrections to scaling due to the $\Or(1/x_g)$ term, we set $x_g=0$ and find that the classical scaling function
\begin{equation}
\Theta_{\infty,d}^{\text{P}}(x_\xi)=\Upsilon_{\infty,d}^{\text{P}}(\infty,x_\xi,0)
\end{equation}
is given by
\begin{eqnarray}
\Theta_{\infty,d}^{\text{P}}(x_\xi)&=&4\pi A_{d+2}\big[\big(\mathcal{X}^{\text{cl}}_{d,\text{as}}\big)^d-x_\xi^d\big]+\Theta_d^{\text{P}}(\mathcal{X}^{\text{cl}}_{d,\text{as}})\nonumber\\ &&\strut +A_d x_\xi^{d-2}\big[(\mathcal{X}^{\text{cl}}_{d,\text{as}})^2-x_\xi^2\big].\qquad
\end{eqnarray}
One easily checks that this equation is consistent with published results \cite{Dan96,Dan98,DDG06} for the classical scaling function. For example, setting the scaled magnetic field $\check{h}=0$ in Eq.~(5.14) of \cite{DDG06}, one sees that the function $2Y_0(x_\xi^2,0)$ of this reference is identical to  $\Theta_{\infty,d}^{\text{P}}(x_\xi)$, as it should.

Until now we restricted ourselves to the bulk disordered phase $\delta\mu \ge 0$. Specializing to the case of $d=3$, we now consider negative and positive values of $\delta\mu$. Since $\nu=1$ at $d=3$, $1/\xi$ behaves linearly in $|\delta\mu|$  for $\delta\mu\to 0-$. Instead of $x_\xi$, we choose the variable
\begin{equation}\label{eq:xdef}
x=-\text{sgn}(\delta\mu) \frac{D}{\xi(-|\delta\mu|)} \text{ for }  d=3,\; \delta\mu\gtreqless 0.
\end{equation}
Because the bulk correlation length is infinite in the ordered phase $\delta\mu>0$, the coefficient of the $x^3$ term of $\Theta_{\infty,3}^{\text{P}}$ for $x<0$ differs from its $x>0$ analog. On the other hand, the result for the scaled inverse finite-size correlation length $\mathcal{X}_{3,\text{as}}^{\text{cl}}(x)$ given in Eq.~\eqref{eq:x3clas} remains valid for $x<0$. It follows that
\begin{align}\label{eq:Thetainf3}
\Theta_{\infty,3}^{\text{P}}(x)=&-\frac{x^3}{12\pi}\,\theta(x)+\frac{x\,\big[\mathcal{X}_{3,\text{as}}^{\text{cl}}(x)\big]^2}{4\pi}-\frac{\big[\mathcal{X}_{3,\text{as}}^{\text{cl}}(x)\big]^3}{6\pi}\nonumber\\ &\strut -\frac{1}{\pi}\Li_3\big[\rme^{-\mathcal{X}_{3,\text{as}}^{\text{cl}}(x)}\big]\nonumber\\&-\frac{1}{\pi}\mathcal{X}_{3,\text{as}}^{\text{cl}}(x)\,\Li_2\big[\rme^{-\mathcal{X}_{3,\text{as}}^{\text{cl}}(x)}\big],\quad(x\gtreqless 0),
\end{align}
which is again consistent with the results of \cite{Dan96} and \cite{Dan98}.

\begin{figure}[htbp]
\begin{center}
\includegraphics[width=0.95\columnwidth]{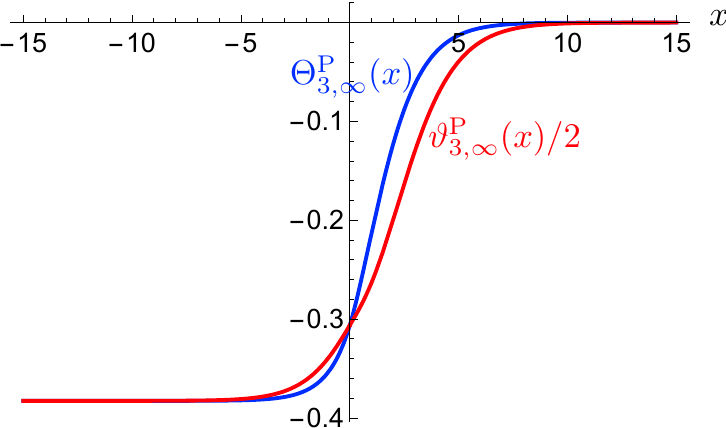}
\caption{The scaling functions $\Theta^{\text{P}}_{3,\infty}(x)$ and $\vartheta^{\text{P}}_{3,\infty}(x)$.}
\label{fig:fig4DR16b}
\end{center}
\end{figure}

The scaling function $\Theta_{\infty,3}^{\text{P}}(x)$ is plotted in Fig.~\ref{fig:fig4DR16b} along with $\frac{1}{2}$ times the associated Casimir force scaling function $ \vartheta_{\infty,3}^{\text{P}}(x)$ one obtains via Eq.~\eqref{eq:Thetavarthetarel}.

The results described in this subsection provide explicit proof of the fact that the finite-size critical behavior which the interacting $n$-component Bose gas  on a  $d$-dimensional strip of finite width $D$ with $2<d<4$ and PBCs exhibits in the limit $n\to\infty$ in the vicinity of the bulk Bose-Einstein transition point is represented by the universality class of the corresponding classical $O(2\infty)$ $\phi^4$ model. Because of the equivalence of the interacting $\infty$-component Bose gas and the imperfect Bose gas, the same statement applies to the latter model, a fact which answers the questions about its  universality class raised in \cite{NJN13}.

It should be clear that the equivalence of these models also holds for ABCs. Consequently, the critical behavior of both models must be described up to quantum corrections by the $O(2\infty)$ $\phi^4$ model with ABCs. The universality class of the latter classical model corresponds (up to a trivial factor of 2 for free energies) to that of the mean spherical model with ABCs studied in \cite{DG09b}.

We refrain from computing the full scaling functions $\Upsilon_{\infty,d}^{\text{A}}(\infty,x_\xi,x_g)$ and $\mathcal{Y}_{\infty,d}^{\text{A}}(\infty,x_\xi,x_g)$ here. However, information about the associated classical scaling functions $\Theta_{\infty,d}^{\text{A}}(x_\xi)=\Upsilon_{\infty,d}^{\text{A}}(\infty,x_\xi,0)$ and $\vartheta_{\infty,d}^{\text{A}}(x_\xi)=\mathcal{Y}_{\infty,d}^{\text{A}}(\infty,x_\xi,0)$ can be inferred from the results of \cite{DG09b}. Note that in their work on the mean spherical model with ABCs the authors of the latter reference allowed for different values $J_\|$ and $J_\perp$ for the ferromagnetic nearest-neighbor bonds parallel and perpendicular to the planes $z=\text{const}$. In the continuum limit this model maps on a $\phi^4$ model whose derivative term $g_{yy}(\nabla_{\bm{y}}\phi)^2+g_{zz}(\partial_z\phi)^2$ involves a diagonal, yet anisotropic metric. This introduces a source of nonuniversality that can be eliminated by an appropriate rescaling of $z$ (see, e.g., p.~15--17 of \cite{DC09}). To obtain the universal scaling functions $\Theta_{\infty,d}^{\text{A}}$ and $\vartheta_{\infty,d}^{\text{A}}$ and the Casimir amplitude from \cite{DG09b}, one can simply set $J_\|=J_\perp$. Specifically, one finds from Eq.~(3.48) of \cite{DG09b} the Casimir amplitude
\begin{equation}
\Delta_{\infty,3}^{\text{A}}=\frac{2}{3}\Im[\Li_2(\rme^{\rmi\pi/3})]-\frac{\zeta(3)}{3\pi},
\end{equation}
whose numerical value $0.549086\ldots$ is also known from \cite{Cha08}. We leave it to the reader to extract from \cite{DG09b} the corresponding predictions for the scaling functions $\Theta_{\infty,3}^{\text{A}}(x)$ and $\vartheta_{\infty,3}^{\text{A}}(x)$.

\subsection{Generalizing the imperfect Bose gas model to allow for nontranslation-invariant boundary conditions}\label{sec:ImBggen}

As we discussed, the interacting $n$-component Bose gas defined by the Hamiltonian~\eqref{eq:Hamn} can be considered for different BCs along the $z$~direction and its $n\to\infty$ limit formulated. For PBC and ABCs, its bulk critical behavior and finite-size critical behavior on a strip of finite width $D$ are the same as those of the imperfect Bose gas. 
This raises the question as to whether appropriate nontranslation-invariant generalizations of the imperfect Bose gas model can be defined that are equivalent to our interacting $n$-component Bose gas with  boundary conditions along the $z$~direction, such as RBCs or DDBCs.

This is in fact possible. Let $\hat{N}_l(z)$ be the number of bosons in layer $z$:
\begin{equation}\label{eq:Nl}
\hat{N}_l(z)=\int \rmd^{d-1}y\,\psih^\dagger(\bm{y},z)\psih(\bm{y},z).
\end{equation}
We now modify the potential energy term in Eq.~\eqref{eq:HIBG} and consider the Hamiltonian
\begin{equation}\label{eq:mHIBG}
\hat{H}_{\text{ImpBG}}=T[\psih^\dagger,\psih]+\frac{a}{2A}\int_0^D\rmd{z}\,[\hat{N}_l(z)]^2.
\end{equation}
We can impose free BCs, such as RBCs or DDBCs, but alternatively also PBCS. 

The equivalence of this modified imperfect Bose gas with the interacting $\infty$-component Bose gas  can be seen as follows. The potential-energy term of $\hat{H}_{\text{ImpBG}}$ yields the contribution
\begin{equation}
S_1[\psi^*,\psi]=\frac{a}{2A}\int_0^{\beta\hbar}\rmd{\tau}\int_0^D\rmd{z}\left[\int\rmd^{d-1}y\,|\psi(\bm{y},z,\tau)|^2\right],
\end{equation}
to the coherent-state action $S_{\text{ImpBG}}=S_0+S_1$ of this model. Using the Hubbard-Stratonovich transformation~\eqref{eq:HStransf}, we can rewrite the exponential $\exp\left(S_1/\hbar\right)$ as a functional integral over a real-valued field $\chi(z,\tau)$ so that the analog of the effective action~\eqref{eq:Seff1} becomes
\begin{eqnarray}\label{eq:SImpBGeff1}
S^{\text{eff}}_{\text{ImpBg}}[\psi^*,\psi,\chi]&=&\int_{\tau,\mathfrak{V}}\bigg[\psi^*\bigg(\hbar\partial_\tau-\mu+\rmi\chi-\frac{\hbar^2}{2m}\nabla^2\bigg)\psi
 \nonumber\\ &&\strut
 +\frac{1}{2a }\chi^2 \bigg].
\end{eqnarray}
We can now introduce a potential $\mathcal{V}(z,\tau)$ by analogy with Eq.~\eqref{eq:Vpotdef},
\begin{equation}\label{eq:VpotdefImpBG}
\mathcal{V}(z,\tau)=\rmi\chi(z,\tau)-\mu,
\end{equation}
perform the functional integral $\int\mathcal{D}[\psi,\psi]$ and exploit the translation invariance along the $\bm{y}$~directions to obtain
\begin{eqnarray}\label{eq:SImpBGeff}
\lefteqn{\ln\int\mathcal{D}[\bm{\psi}^*,\bm{\psi}]\,\rme^{-S_{\text{eff}}[\bm{\psi}^*,\bm{\psi},\chi]/\hbar}+\text{const}}&&\nonumber\\
&=&-A\int_{\tau,z}\bigg\{\pint\Big\langle z,\tau\Big|\ln\Big[\hbar\partial_\tau+\mathcal{V}\nonumber\\&&\strut+\frac{\hbar^2}{2m}(p^2-\partial_z^2)\Big]\Big|z,\tau\Big\rangle
-\frac{ 1}{2 a}(\mu+\mathcal{V})^2\bigg\}/\hbar.\qquad
\end{eqnarray}

Since the right-hand side is proportional to hyperarea $A$,  the remaining functional integral $\int\mathcal{D}[\chi]$ can be calculated in the thermodynamic limit $A\to\infty$ by evaluating its integrand at the stationary point. The result shows that our modified imperfect Bose gas model defined by Eqs.~\eqref{eq:mHIBG} and \eqref{eq:Nl}  is in one-to-one correspondence with the $n\to\infty$ limit of the interacting Bose gas, if we identify the interaction strength $a$ with the coupling constant $\ub$ of the latter. 

\subsection{Results for $\bm{d=3}$ with Dirichlet-Dirichlet boundary conditions}

From our general considerations based on the coherent-state functional-integral approach in Sec.~\ref{sec:models} and our analysis in Sec.~\ref{sec:intBG} it should be clear that the asymptotic large-scale finite-size behavior of our ${n=\infty}$~Bose film model with free BCs in the scaling regime of  the bulk Bose-Einstein transition point is described by the  classical $O(2\infty)$ $\phi^4$ model. Unfortunately, exact analytic solutions for free BCs  are  neither known for the self-consistent potential $\mathcal{V}_{*}(z)$ nor for the eigenvalues $\ev_\nu$ and eigenfunctions $\ef_\nu$ even if quantum corrections are neglected. Of particular interest is the ${(d=3)}$-dimensional case on which we now focus. For it, a number of exact analytic results have been obtained for the classical theory with DDBCs \cite{BM77a,BM77c,DR14,RD15a,RD15b,DGHHRS12,DGHHRS14,DR17}, which  are known to apply to this theory with free BCs at ${d=3}$  asymptotically in the large length scale limit. Using a combination of techniques such as direct solutions of the self-consistent equations \cite{DR14}, short-distance and boundary-operator expansions \cite {DGHHRS14}, trace formulas \cite{RD15b}, inverse scattering methods for the semi-infinite case ${D=\infty}$ and matched semiclassical expansions for ${D<\infty}$ \cite{RD15a}, exact analytic results for several  series expansion coefficients of the self-consistent potential $v_*(z)$ and for the asymptotic ${x\to\infty}$ behaviors of the eigenvalues $\ev_\nu^{\text{DD}}$, eigenfunctions $\ef^{\text{DD}}_\nu$, and the classical scaling functions of the residual free energy and the Casimir force have been determined. However, the computation of  these scaling functions for all values $x\gtreqless 0$ of the scaling variable $x$ introduced in Eq.~\eqref{eq:xdef} required the use of numerical methods \cite{DGHHRS12,DGHHRS14,DGHHRS15}.

In order to solve the self-consistent Schr\"odinger equation numerically, it must be discretized. We do this in the same manner as in the treatment of the classical model called A in \cite{DGHHRS12} and \cite{DGHHRS14}, i.e., we discretize only in the $z$~direction, keeping the $\bm{y}$~coordinates continuous. Let $a_z$ be the corresponding lattice spacing. Then, the discretized system consists of $N_z=L/a_z$ layers located at
\begin{equation}
z_j=(j-1/2)a_z,\,j=1,\dotsc,N_z.
\end{equation}
For convenience we set $a_z=1$. The discrete analog of the Schr\"odinger Eq.~\eqref{eq:Seq} is the eigenvalue equation
\begin{equation}
\sum_{j=1}^{N_z}H_{j,j'}\ef_{\nu,j'}=\ev_\nu\ef_{\nu,j}
\end{equation}
for the matrix 
\begin{equation}\label{eq:Hmatrix}
\mat{H}=-\mat{D}^2+\mat{v}
\end{equation}
with the tridiagonal discrete Laplacian (for DDBCs)
\begin{equation}\label{eq:D2def}
\mat{D}^2=
\begin{pmatrix}
-2&\phantom{+}1&&\\
\phantom{+}1&\ddots&\ddots&\\
&\ddots&\ddots&\phantom{+}1\\[0.1em]
&&\phantom{+}1&-2
\end{pmatrix}
\end{equation}
and the diagonal potential matrix $\mat{v}=\text{diag}(v_1,\dotsc,v_{N})$.

Since the bulk limit $N_z\to\infty$ of these equations is independent of the BC, we can choose PBCs to study it. The spectrum $\{\ev_\nu\}$  becomes dense as $N_z\to\infty$. Because of the modified (lattice) dispersion relation, the bulk eigenvalues $\ev^{\text{dct}}_{\text{b}}(k)$ of our discretized model are given by
\begin{equation}\label{eq:epsmclat}
\ev^{\text{dct}}_{\text{b}}(k)=4\sin^2(k/2)+v_{\text{b},*}\,,\quad 0\le k\le \pi,
\end{equation}
rather than by $k^2+v_{\text{b},*}$. 
The changes this implies for our results for the bulk grand potential, the bulk self-consistency equation, and the critical value $\mu_{\text{c}}(T,\ub)$ of the chemical potential are equivalent to the replacement of the pair correlation function $g_{\text{b},d}(\bm{x};\lamth,\xi)$ by its analog for our discretized model, namely,
\begin{align}\label{eq:gbdctdef}
&g^{\text{dct}}_{\text{b},d}(\bm{x},\lamth,\xi)\nonumber\\ &\equiv\int_0^\pi\frac{\rmd{k}}{\pi}\pint\frac{\rme^{\rmi\bm{x}\cdot(\bm{p},k)}}{\rme^{\lamth^2[p^2+\xi^{-2}+4\sin^2(k/2)]/4\pi}-1}.
\end{align}
Thus the analogs of Eqs.~\eqref{eq:inftybd<}, \eqref{eq:inftybd>}, and \eqref{eq:muc} for our discrete model can be written as
\begin{equation}\label{eq:inftybopdct}
\varphi^{\text{dct},<}_{\infty,\text{b},d}(T,\mu,\ub)=-\lamth^2\,g^{\text{dct}}_{\text{b},d+2}(\bm{0};\lamth,\xi)-\frac{\beta}{2\ub}(\mu+\mathcal{V}_{\text{b},*})^2,
\end{equation}
\begin{equation}
\varphi^{\text{dct},>}_{\infty,\text{b},d}(T,\mu,\ub)=-\lamth^2\,g^{\text{dct}}_{\text{b},d+2}(\bm{0};\lamth,0)-\frac{\beta}{2\ub}\mu^2,
\end{equation}
\begin{equation}
\mathcal{V}_{\text{b},*}+\mu=\ub\,g^{\text{dct}}_{\text{b},d}(\bm{0};\lamth,\xi),
\end{equation}
and
\begin{equation}
\mu^{\text{dct}}_{\text{c}}(T,\ub)=\ub  g^{\text{dct}}_{\text{b},d}(\bm{0};\lamth,0),
\end{equation}
respectively. Note that these bulk quantities also depend on the discretization length $a_z$, which we have set to $1$. 

The classical limit $\lamth\to 0$ of $g^{\text{dct}}_{\text{b},d}(\bm{x};\lamth,\xi)$ is related to the  correlation function
\begin{equation}\label{eq:gbcldctdef}
g^{\text{cl,dct}}_{\text{b},d}(\bm{x};\xi)\equiv \int_0^\pi\frac{\rmd{k}}{\pi}\pint\frac{\rme^{\rmi\bm{x}(\bm{p},k)}}{p^2+4\sin^2(k/2)+\xi^{-2}}
\end{equation}
of the correspondingly discretized classical $O(\infty)$ $\phi^4$ model. By analogy with Eq.~\eqref{eq:gbgbcl}, we have
\begin{equation}\label{eq:gbgbcldct}
\lim_{\lamth\to 0}\frac{\lamth^2}{2\pi}\,g^{\text{dct}}_{\text{b},d}(\bm{x};\lamth,\xi)=2g^{\text{cl,dct}}_{\text{b},d}(\bm{x};\xi).
\end{equation}

The integrals in Eq.~\eqref{eq:gbcldctdef} can be computed using dimensional regularization for the $\bm{p}$~integral. One obtains \cite{DGHHRS14,fn7}
\begin{equation}\label{eq:gbdcrtres}
g^{\text{dct}}_{\text{b},d}(\bm{x},\lamth,\xi)=-A_{d-1}\xi^{3-d}{}_2F_{1}(\tfrac{1}{2},\tfrac{3-d}{2};1;-4\xi^2).
\end{equation}
 
This result can be exploited in a straightforward fashion to  derive the analog of Eq.~\eqref{eq:IdBGcf}. Upon considering the $\lamth\to 0$ limit of the self-consistency equation and substituting the large-$\xi$ expansion of $ g^{\text{dct}}_{\text{b},d}(\bm{x},\lamth,\xi)$ into it, one finds $\xi\asprop |\mu-\mu^{\text{dct}}_{\text{c}}|^{-1/(d-2)}$, where the proportionality constant differs from the one implied by Eq.~\eqref{eq:IdBGcf} because of the different dispersion relation of the discrete model. Moreover, upon introducing a variable $\tb$ such that 
\begin{equation}
\beta\mu=-\lamth^2\tb/4\pi
\end{equation}
 and expressing $\ub=\hbar^2\lamth^2g/24\pi m$  in terms of the coupling constant $g$ introduced in Eq.~\eqref{eq:gdef} and $\mathcal{V}_{\text{b}}$ in terms of $v_{\text{b}}$, we can take the limit $\lamth\to 0$ of the bulk grand potential at fixed $\tb$ and $g$ to obtain the associated classical bulk free energy density. One obtains
\begin{align}\label{eq:fblim}
&\lim_{\substack{\lambda\to 0\\ \tb,g}}\varphi^{\text{dct}}_{\infty,\text{b},d}(T,\mu,\ub)\equiv f^{\text{dct}}_{\infty,\text{b},d}(\tau,g)\nonumber\\
&=-4\pi A_{d+1\,}g^{\text{cl,dct}}_{\text{b},d+2}\big(\bm{0};v_{\text{b}}^{-1/2}\big)-\frac{3}{g}\left(\tau+\tb_{\text{c}}-v_{\text{b}}\right)^2,
\end{align}
where 
\begin{equation}
\tau=\tb-\tb_{\text{c}}=4\pi\lamth^{-2}\beta\,(\mu_{\text{c}}-\mu)
\end{equation}
is the deviation of $\tb$ from its bulk critical value $\tb_{\text{c}}$ .

Evaluated at the stationary point $v_{\text{b},*}$, the  result is the bulk free energy density of the classical model with the coherent-state action~\eqref{eq:Scl} in the limit $n\to\infty$. Hence, the correspondence of the classical limit of our ${n=\infty}$ Bose model  with the $O(2\infty)$ $\phi^4$ model carries over to the discretized versions of these models. We have explicitly verified this here only for the bulk grand potential, but it should be obvious that the correspondence of $\varphi_{\infty,d}^{\text{DD}}$ in the classical limit with the reduced free energy per layer, $f^{\text{DD}}_{\infty,d}(\tb,\ub,D)$, of the $O(2\infty)$ $\phi^4$ also holds for the discretized versions.
However, to understand in detail that, and how, the results for $d=3$ of \cite{DGHHRS12}, \cite{DGHHRS14}, and \cite{DGHHRS15} are related to the scaling behavior of our Bose model, a few explanatory remarks will be helpful.

(i) One cannot simply set ${d=3}$ in the classical theory because the dimensionally regularized functions $g^{\text{dct}}_{\text{b},d}(\bm{0},\lamth,\xi)$ and $g^{\text{dct}}_{\text{b},d+2}(\bm{0},\lamth,\xi)$ have simple poles at ${d=3}$ corresponding to UV singularities. The Laurent expansions of these functions about ${d=3}$ are known from \cite{DrFSchmidt}, \cite{DGHHRS12}, and \cite{DGHHRS14}. For our purposes it is sufficient to know that the differences of the first function and its value at the bulk transition point, and that of the second function and its Taylor series expansion to first order in $\xi^{-2}$,
\begin{equation}
S_d^{(1)}(\xi)= \sum_{j=0}^1\xi^{-2j}\big[\partial^j_rg^{\text{dct}}_{\text{b},d+2}(\bm{0},\lamth,r^{-1/2})\big]_{r=0},
\end{equation}
have finite ${d\to3}$ limits. We have
\begin{equation}\label{eq:gbd3reg}
\lim_{d\to 3}[g^{\text{dct}}_{\text{b},d}(\bm{0},\lamth,\xi)-g^{\text{dct}}_{\text{b},d}(\bm{0},\lamth,\infty)]=-\frac{\arsinh(1/2\xi)}{2\pi}
\end{equation}
and 
\begin{align}\label{eq;gbd5reg}
&\lim_{d\to 3}[g^{\text{dct}}_{\text{b},d+2}(\bm{0},\lamth,\xi)-S^{(1)}_d(\xi)]\nonumber \\ &=\frac{1}{16\pi^2\xi^2}\Big[(2+4\xi^2)\arsinh(1/2\xi)-\sqrt{1+4\xi^2}\Big].
\end{align}

(ii) The first difference is encountered automatically if one subtracts from the classical bulk self-consistency equation its analog at the bulk transition point. To make the bulk free energy $f^{\text{dct}} _{\infty,\text{b},d}$ UV~finite, we can follow \cite{DGHHRS14} and  subtract from it  its Taylor expansion to first order in $\tau$,
\begin{equation}
S(\tau,g)=f_{\infty,\text{b},d}^{\text{dct}}(0,g)-A_{d-1}\tau,
\end{equation}
defining the renormalized bulk free energy
\begin{equation}\label{eq:fbren}
f_{\infty,\text{b},d}^{\text{dct,ren}}(\tau,g)\equiv f_{\infty,\text{b},d}^{\text{dct}}(\tau,g)-S_d(\tau,g).
\end{equation}
 Its   limit for ${d\to3}$  is twice the expression given in Eq.~(4.16) of \cite{DGHHRS14}, namely,
\begin{align}\label{eq:fbrenres}
f_{\infty,\text{b}}^{\text{ren}}(\tau,g )&=\frac{1}{4\pi\xi^2} \sqrt{  4\,\xi^2+  1}-\frac{2+  1/\xi^2}{2\pi}\arsinh(1/2\xi)\nonumber\\
 & -\frac{3}{g }(\tau- 1/\xi^2)^{2},
\end{align}
where $\xi$, the bulk correlation length, satisfies
\begin{equation}\label{eq:Vb(dt,g)}
 \xi^{-2}=\begin{cases}
\tau-\frac{g }{12\pi}\arsinh(1/2\xi) & \text{for }\tau>0,\\
0 &\text{for } \tau\leq 0.
\end{cases}
\end{equation}

(iii) By analogy with Eqs.~\eqref{eq:fblim} and \eqref{eq:fbren}, we can take the classical limit of the layer grand potential at finite $D$ to obtain the reduced layer free energy
\begin{equation}
f_{\infty,d}^{\text{dct}}(\tau,g,D)\equiv \lim_{\substack{\lambda\to 0\\ \tb,g}}\varphi^{\text{dct}}_{\infty,d}(T,\mu,\ub,D)
\end{equation}
and introduce the  renormalized quantity
\begin{equation}\label{eq:fDren}
f_{\infty,d}^{\text{dct,ren}}(\tau,g,D)\equiv f_{\infty,d}^{\text{dct}}(\tau,g,D)-D\,S_d(\tau,g). 
\end{equation}
Its UV-finite limit at $d=3$ is twice the result given in Eq.~(4.15) of \cite{DGHHRS14}. We do not give it here since we  are not going to use it in the following.

(iv) Once the classical limit of our Bose model has been taken to eliminate the corrections to scaling due to quantum effects, the analyses of the corrections to scaling performed in \cite{DGHHRS12} and \cite{DGHHRS14} fully apply to the remaining classical ones of our Bose model. In particular, one can eliminate corrections to scaling by taken the limit $g\to\infty$. To this end, one defines at ${d=3}$ a linear scaling variable 
\begin{equation}\label{eq:tdef}
t=24\pi\tau/g
 \end{equation}
 in which the amplitude of the correlation length $\xi$ for $\tau >0$ ($\mu<\mu_{\text{c}}$) has been absorbed. Upon adding to $f_{\infty,3}^{\text{dct,ren}}(\tau,g,D)$ the term $D 3\tau^2/g$, one can perform the limit $g\to\infty$ to obtain the finite $t$-dependent layer free energy
\begin{equation}\label{eq:fDtren}
f_{\infty,3}^{\text{dct,ren}}(t,D)=\frac{1}{4\pi}\Tr\!\left[\mat{H} \left(1+t-\ln \mat{H} \right)\right]-\frac{t D}{2\pi}\,.
\end{equation}
Furthermore, the self-consistency equation, the bulk free energy, and the bulk correlation length at $d=3$ simplify in this  limit ${g\to\infty}$ to
 \begin{equation}\label{eq:sct}
t=\langle z| \ln\mat{H} |z\rangle,
\end{equation}
\begin{equation}\label{eq:fbtren}
f_{\infty,\text{b},3}^{\text{dct,ren}}(t)=\frac{1}{2\pi}
\begin{cases}
\sinh (t)-t &\text{for } t>0,\\
0 &\text{for } t\le 0,
\end{cases}
\end{equation}
and
\begin{equation}
\xi=\begin{cases}
[2\sinh(t/2)]^{-1} &\text{for } t>0,\\
\infty &\text{for } t\le 0.
\end{cases}
\end{equation}
These equations were used in \cite{DGHHRS12} and \cite{DGHHRS14} to determine the classical scaling functions of the residual free energy and the Casimir force quite accurately by numerical means. The corresponding finite-$g$ equations were also studied there and their consistency with the ${g=\infty}$ results verified.

The upshot of these considerations is that the numerical results of \cite{DGHHRS12,DGHHRS14,DGHHRS15} for the self-consistent potential give us directly the potential $v(z)$ up to the ignored exponentially small quantum corrections, whereas those  for the bulk, layer, and residual free energies and the Casimir force must be multiplied by a factor of 2 to give us their analogs for the Bose gas in the scaling regime near the bulk critical point up to exponentially small quantum corrections. For example, the plot of the critical potential shown in Fig.~3 of \cite{DGHHRS14} directly applies to our Bose model, and the scaling functions $\Theta(x)$ and $\vartheta(x)$  displayed in Fig.~4 of this reference correspond to the functions $\Theta_{\infty,3}^{\text{DD}}(x)/2$ and $\vartheta_{\infty,3}^{\text{DD}}(x)/2$.

The exact analytic results for the $O(\infty)$ $\phi^4$ model obtained or reported in \cite{DGHHRS12,DGHHRS14,DR14,RD15a,RD15b} and \cite{DR17} can be translated to the Bose gas case in a similar fashion. We give a few examples. First, the potential $v(z)$, which is symmetric with respect to reflections about the midplane $z=D/2$, i.e.,  $v(z)=v(D-z)$, behaves asymptotically as  
\begin{equation}\label{eq:v3ddcrit}
v_{d=3}(z,t,D)\mathop{=}_{\lamth\ll z\ll 1/|t|}-\frac{1}{4z^2}+\frac{4t}{\pi^2 z}+\frac{56\zeta(3)}{\pi^4}\,t^2+\ldots.
\end{equation}

Second, at bulk criticality $t=0$, the second (``far'') boundary plane produces a leading correction $\asprop D^{-3}$ so that 
\begin{equation}\label{eq:vDD30}
v_{d=3}(z,0,D)\mathop{=}_{\lamth\ll z\ll D}\frac{-1}{4z^2}\left[1-\frac{512}{\pi}\,\Delta_{\infty,3}^{\text{DD}} \frac{z^3}{D^3}\right].
\end{equation}
Here $\Delta_{\infty,3}^{\text{DD}} $ denotes the Casimir amplitude
\begin{equation}
\Delta_{\infty,3}^{\text{DD}} =\Theta_{\infty,3}^{\text{DD}}(0)= -2\times 0.01077340685024782(1),
\end{equation}
whose quoted numerical value is taken from \cite{DGHHRS14}. Moreover, the scattering data that are equivalent to the potential $v_{d=3}(z;t,{D=\infty})$ for the semi-infinite case ${D=\infty}$ are known in closed analytical form; they can be found in  Eqs.~(4.54), (4.64), and (4.66)--(4.68) of \cite{RD15b}.

Third, also known from \cite{RD15a} are the leading singular behaviors of $\Theta_{\infty,3}^{\text{DD}}(x)$ and
\begin{equation}
\vartheta_{\infty,3}^{\text{DD}}(x)=2\,\Theta_{\infty,3}^{\text{DD}}(x)+x\frac{\rmd}{\rmd x}\Theta_{\infty,3}^{\text{DD}}(x)
\end{equation}
as  $x\to 0\pm$. One has 
\begin{align}\label{eq:Thetacrit}
\Theta^{\text{DD}}_{\infty,3}(x)\mathop{=}_{x\to 0\pm}& \Delta_{\infty,3}^{\text{DD}}-\left[\Delta A_0^{(\rm{s})}+\frac{x}{24\pi}\right]2x^2\,\Htheta(x)\nonumber\\&\strut +\frac{1}{\pi^3}\,x^2\ln|x|+\sum_{j=1}^3\Theta_j x^j+\text{o}(x^3),
\end{align}
where 
\begin{align}\label{eq:DelA0anexp}
\Delta A^{(\mathrm{s})}_0&=\frac{1}{8\pi}\int_0^\infty\rmd{u}\,\frac{\coth(u)-1/u}{u^2+\pi^2/4}\nonumber\\
&= 0.01888264398\ldots.
 \end{align}
is a universal amplitude difference while the $\Theta_j$ are coefficients of regular contributions. The result yields the exact value
\begin{equation}
\big[\vartheta_{\infty,3}^{\text{DD}}\big]''(0)=-2/\pi^3.
\end{equation}

Finally, we mention that three terms of the asymptotic expansions of the functions $\Theta^{\text{DD}}_{\infty,3}(x)$ and $\vartheta^{\text{DD}}_{\infty,3}(x)$ for $x\to-\infty$ are known. For the first function, it reads as \cite{RD15a}
\begin{eqnarray}\label{eq:Thetalargemx}
\Theta^{\text{DD}}_{\infty,3}(x)&\mathop{=}\limits_{x\to-\infty}&\frac{-\zeta(3)}{8\pi}\bigg[1+\frac{2}{|x|}\bigg(\gamma_{\text{E}}-\frac{1}{2}-\frac{\zeta'(3)}{\zeta(3)}+\ln\frac{4|x|}{\pi}\bigg)\nonumber\\ &&\strut+\text{o}\Big(\frac{1}{|x|}\Big)\bigg].
\end{eqnarray}

\section{Summary and conclusions\label{sec:sumconc}}

We investigated fluctuation-induced forces in Bose gases confined to strips of thickness $D$ near their Bose-Einstein bulk condensation point. Both the cases of ideal and nonideal Bose gases have been considered. For convenience, we present here a brief overview of our results, putting them in context with previously published ones and referencing our most important equations. 
We consider separately the parts dealing with ideal and nonideal Bose gases.

\emph{(i) Ideal Bose~gas case.} In \cite{MZ06} the residual grand potential and the Casimir force have been determined at ${d=3}$ in the form of double series for the cases of PBCs, DDBCs, and NNBCs along the finite direction. There, it has also been demonstrated that quantum effects contribute exponentially small corrections in the scaling regime near the bulk transition point. 
The double series that follow from the results of this reference for the scaling functions $\Upsilon_d^{\text{P}}(x_\xi,x_\lambda)$ and $\Upsilon_d^{\text{DD}}(x_\xi,x_\lambda)=\Upsilon_d^{\text{NN}}(x_\xi,x_\lambda)$ are listed in Eqs.~\eqref{eq:Ydapres} and \eqref{eq:UpsDDNN}. We have confirmed them by means of  alternative derivations. 

Furthermore, we have generalized the above-mentioned results to ABCs, DNBCs, and RBCs. The analog of the double series~\eqref{eq:Ydapres} and \eqref{eq:UpsDDNN} for $\Upsilon_d^{\text{A}}(x_\xi,x_\lambda)$ is given in Eq.~\eqref{eq:Ydapres}. Our result for $\Upsilon_d^{\text{DN}}(x_\xi,x_\lambda)$  is covered by the one for $\Upsilon_d^{\text{R}}(x_\xi,x_\lambda)$ given in Eqs.~\eqref{eq:varphiRres} and \eqref{eq:UpsRresd} as the special case $(\scc_1,\scc_2)=(\infty,0)$. Likewise, the latter equations apply to $\Upsilon_d^{\text{DD}}(x_\xi,x_\lambda)$ and $\Upsilon_d^{\text{NN}}(x_\xi,x_\lambda)$ if $(\scc_1,\scc_2)=(\infty,\infty)$ and $(\scc_1,\scc_2)=(0,0)$, respectively. The representation~\eqref{eq:UpsRresd}  of $\Upsilon_d^{\text{R}}$ provides an explicit decomposition into a leading classical contribution and a sum $\sum_{\rho=1}^\infty$ of quantum corrections depending on the Matsubara frequencies $\omega_\rho$. These quantum corrections decay exponentially with $D$ on length scales that decrease as $\rho$ increases. At bulk criticality,  the leading quantum correction decay $\asprop \rme^{-D/l_1}$, where the length scale $l_1\propto \lamth$ is half as big  for free BC such as DDBs, NNBCs, DNBCs, and RBCs  as for PBC.
 
If the quantum corrections in the results of  \cite{MZ06} are dropped, the double series for $\Upsilon_d^{\text{P}}(x_\xi,x_\lambda)$ and $\Upsilon_d^{\text{DD}}(x_\xi,x_\lambda)=\Upsilon_d^{\text{NN}}(x_\xi,x_\lambda)$ must reduce to series for the classical scaling functions of the free massive  $O(2)$ theory. This was pointed out and verified in \cite{GD06a}. We have explicitly shown that the same holds true for ABCs, DNBCs, and RBCs. 

Summing the series of the  scaling functions $\Theta_3^{\text{BC}}$ with $\text{BC}=\text{P, A, DD, NN,}$ and $\text{DN}$ of the classical free $O(2)$ theory, we have derived the closed exact analytic expressions for $\Theta^{\text{BC}}_3(x_\xi)$ and $\vartheta^{\text{BC}}_3(x_\xi)$ given in Eqs.~\eqref{eq:Theta3per}--\eqref{eq:Theta3DN} and \eqref{eq:vartheta3per}--\eqref{eq:vartheta3DN}, respectively.

\emph{(i) Nonideal Bose~gas case.} 
We have considered two distinct, but related, models for nonideal Bose gases on a strip subject to different BC along the finite direction: the so-called imperfect Bose gas \cite{Dav72,NJN13} and an $n$-component generalization of a standard Bose model with short-range interactions. The first one, defined for PBC, was investigated in \cite{NJN13}. There,  the critical Casimir force was computed right at the Bose-Einstein bulk transition point in ${d=3}$ dimensions, but the full scaling functions near this transition not determined. The amplitude of this force turned out to have twice the value it has for the mean-spherical model  \cite{Dan96,Dan98}, which prompted the authors to raise the question  as to which universality class applies to  the fluctuation-induced forces of the imperfect Bose gas.

We have shown that the imperfect Bose gas with PBC corresponds to the $n\to\infty$ limit of our $n$-component Bose model with short-range interactions.  It follows from the general arguments discussed in Sec.~\ref{sec:models} that the bulk critical behavior  and the finite-size critical behavior of the latter model on the strip are represented by the corresponding classical $O(2n)$ $\phi^4$ model. As a consequence, the critical Casimir forces near the Bose-Einstein bulk transition point of the imperfect Bose gas with PBCs  must be representative of the universality class of the $O(2n)$ $\phi^4$ model in the limit $n\to\infty$. Since the $O(n)$ $\phi^4$ model with PBCs in the limit $n\to\infty$ belongs to the same universality class as the mean spherical model, the bulk critical and finite-size critical behaviors of the imperfect Bose gas near the bulk transition point are represented by the latter model up to a trivial factor of $2$ in the free energies and the Casimir force. We have explicitly verified this by computing the scaling functions $\Theta_{\infty,3}^{\text{P}}(x)$ and $\vartheta^{\text{P}}_{\infty,3}(x)$ from the $\infty$-component Bose model, proving that they comply with the results of \cite{Dan96,Dan98} for the mean spherical model (up to the mentioned trivial factor of $2$). Our exact analytic result for $\Theta_{\infty,3}^{\text{P}}(x)$ is given in Eq.~\eqref{eq:Thetainf3}, and plots of the functions $\Theta_{\infty,3}^{\text{P}}(x)$ and $\vartheta^{\text{P}}_{\infty,3}(x)$ are displayed in Fig.~\ref{fig:fig4DR16b}.

The equivalence of the imperfect Bose model and the $\infty$-component Bose model suggests a generalization of the former for other BC, namely, free BC. We have introduced such a generalized imperfect Bose model with free BC along the $z$~direction in Sec.~\ref{sec:ImBggen}. Its bulk critical and finite-size critical behaviors near the bulk critical point are represented by the corresponding $O(2\infty)$ $\phi^4$ with free BC. At ${d=3}$, where RBCs with subcritical enhancement variables $c_1>0$ and $c_2>0$ turn into DDBCs in the large length~scale limit, one can therefore exploit the known exact results for the $O(\infty)$ $\phi^4$ model \cite{DGHHRS12,DGHHRS14,DGHHRS15,DR14,RD15a,RD15b,DR17} to obtain exact information about the corresponding scaling functions $\Theta_{\infty,3}^{\text{DD}}(x)$ and $\vartheta^{\text{DD}}_{\infty,3}(x)$. A number of exact analytic properties, such as the near-boundary behavior of the self-consistent potential and the limiting behaviors of the scaling functions for $x\to-\infty$, are presented in Eqs.~\eqref{eq:v3ddcrit}, \eqref{eq:vDD30}, and \eqref{eq:Thetacrit}--\eqref{eq:Thetalargemx}. In order to benefit also from the numerical results of \cite{DGHHRS12} and \cite{DGHHRS14}, we have generalized the discretization scheme of the discretized $O(\infty)$ $\phi^4$ model called A in these references to the $n$-component Bose gas and verified that the numerically computed scaling functions correspond to one-half of  those of the $\infty$-component Bose gas, namely, $\Theta^{\text{DD}}_{\infty,3}(x)/2$ and $\vartheta^{\text{DD}}_{\infty,3}(x)/2$.

The primary focus of our investigations here has been put on fluctuation-induced forces in the scaling regime of the bulk critical point. Accordingly, we have assumed throughout this paper that  the strip thickness $D$ is much larger than the thermal de Broglie wavelength $\lamth$. However, as $T$ decreases at fixed given $D$, the thermal length $\lamth$ ultimately becomes much larger than the thickness $D$. This suggests complementary studies of the asymptotic regime $\lamth\gg D$ both for  ideal and interacting Bose gases.  Rather than embarking on such a study, we  restrict ourselves here to a few remarks.

An investigation of fluctuation-induced forces in the asymptotic regime $\lamth\gg D$ was made for the imperfect Bose gas on a strip with PBCs and $2<d<4$  in a recent paper  \cite{JNS16}. In this case one can restrict oneself to the  ${k^{\text{P}}=0}$ mode of the operator $-\partial_z^2$ because the modes with eigenvalues $(k_\nu^{\text{P}})^2=(2\pi\nu/D)^2$ give subleading (exponentially decaying) corrections. The essence of this approximation (used in \cite{JNS16}) is easily understood in the language of the coherent-state representation of the model. Written in terms of the (${k=0}$)~field component $\Psi(\bm{y},\tau)\equiv D^{-1/2}\int_0^D\rmd{z}\,\psi(\bm{y},z,\tau)$, the effective action $S_{\text{eff}}[\Psi^*,\Psi]$ one obtains upon discarding the remaining ${k^{\text{P}}>0}$ contributions to $\psi(\bm{y},z,\tau)$ describes a (${d-1}$)-dimensional interacting Bose field theory with a coupling constant $\ub/D$, where $\ub$ is the interaction constant of the $d$-dimensional theory \cite{fn8,BDZ08}. Reference~\cite{JNS16} finds three regions of distinct asymptotic behaviors distinguished by whether $(|\mu|/\ub)/\lamth^2D\gg 1$ or $\ll 1$ and the sign of $\mu$. Whether and to what extent these findings might carry over to interacting Bose gases on a three-dimensional strip with PBCs is not clear to us for two reasons. The first is that the effective $(d-1)$-dimensional interacting field theory that results upon making the replacement $\psi(\bm{y},z,\tau)\to\Psi(\bm{y},\tau)$ appears to require at low temperatures a more sophisticated treatment  than the Hartree-type approximation to which the use of the imperfect Bose~gas model corresponds.  Renormalization group analyses of the low-temperature behavior of interacting Bose gases such as  \cite{PCDS04} and \cite{FW09} indicate this. (For further literature on such approaches, see the references of these papers and those of \cite{BDZ08}.)  Furthermore, for ${d=3}$, the corresponding effective two-dimensional interacting Bose gas has a low-temperature phase with quasi long-range order, which the imperfect Bose gas misses. 


\begin{acknowledgments}
We are grateful to M.\ Napi{\'o}rkowski for informing  us about his work on the imperfect Bose gas and informative correspondence, and also to W.\ Zwerger for his helpful comments.
\end{acknowledgments}

\appendix
\section{Scaling functions of the ideal Bose gas\label{app:idBG}}
To compute $\Upsilon_d^{\text{P}}$ we use the fact that $\varphi_{\text{s},d}^{\text{P}}=0$, introduce the variables $x_\lambda\equiv D/\lamth$ and $x_\xi\equiv D/\xi$, and set $D=1$ so that $\Upsilon_d^{\text{BC}}(x_\lambda,x_\xi)=\varphi_{\text{res},d}^{\text{BC}}(T,\mu,1)$. After an integration by parts, one arrives at
\begin{eqnarray}\label{eq:Upsperd}
\lefteqn{\Upsilon_d^{\text{P}}(x_\lambda,x_\xi)=-\frac{K_{d-1}}{(d-1)2\pi x_\lambda^2}}&&\nonumber\\ &&\times\left(\sum_{k/2\pi \in\mathbb{Z}}-\int_{-\infty}^\infty\frac{\rmd{k}}{2\pi}\right) \int_0^\infty\frac{\rmd{p}\,p^d}{\exp\Big[\frac{k^2+p^2+x_\xi^2}{4\pi x_\lambda^2}\Big]-1}\nonumber\\
&=&-x_\lambda^{d-1}\left(\sum_{k/2\pi\in\mathbb{Z}}-\int_{-\infty}^\infty\frac{\rmd{k}}{2\pi}\right)\text{Li}_{\frac{d+1}{2}}\bigg(\rme^{-\frac{k^2+x_\xi^2}{4\pi x_\lambda^2}}\bigg).\quad
\end{eqnarray}
The subtracted term involving the $k$~integral is the bulk term 
\begin{equation}
\varphi_{\text{b},d}(T,\mu)=-\lamth^{-d}\,\text{Li}_{\frac{d+2}{2}}(\mathfrak{z}),\;\;\;\mathfrak{z}=\rme^{\beta\mu}=\rme^{-\frac{1}{4\pi}x_\xi^2/x_\lambda^2},
\end{equation}
where $\mathfrak{z}$ is the fugacity.

We now substitute  the series representation $\Li_\nu(x)=\sum_{s=1}^\infty s^{-\nu} x^s$ for the polylogarithm in Eq.~\eqref{eq:Upsperd} and use Poisson's summation formula~\eqref{eq:Psumform} with ${D=1}$ for $f(k)=\exp(-sk^2/4\pi x_\lambda^2)$. This gives  the result stated in Eq.~\eqref{eq:Ydperres}.

To compute the scaling function $\Upsilon^{\text{A}}_d$, one can use the generalized Poisson identity for theta functions
\begin{equation}
\sum_{\nu=-\infty}^\infty\rme^{-t(\nu+a)^2}=(\pi/t)^{1/2}\sum_{j=-\infty}^\infty\rme^{-\pi^2j^2/t}\cos(2\pi aj)
\end{equation}
with $t=\pi s/x\lambda^2$ and $a=1/2$. A straightforward calculation then yields Eq.~\eqref{eq:Ydapres}.

For  DDBCs, NNBCs, and DNBCs a more general form of Poisson's formula  involving the cosine transform
\begin{equation}
f_{\text{cos}}(x)\equiv \int_0^\infty\rmd{\nu}\,f(\nu)\cos(x\nu).
\end{equation}
can be used, namely,
\begin{equation}\label{eq:Poissum}
\sum_{\nu=1}^\infty f(\nu)=-\frac{1}{2}f(0)+f_{\text{cos}}(0)+2\sum_{j=1}^\infty f_{\text{cos}}(2\pi j),
\end{equation}
which holds for non-negative, continuous,  decreasing, and Riemann integrable functions  $f$ on $[0,\infty)$ \cite{Apo57}.

It yields the Jacobi identity 
\begin{equation}\label{eq:Jacobid}
\sum_{\nu=1}^\infty\rme^{-s(\pi\nu)^2/4\pi x_\lambda^2}=\frac{x_\lambda}{s^{1/2}}-\frac{1}{2}+\frac{2x_\lambda}{s^{1/2}}\sum_{j=1}^\infty\rme^{-4\pi x_\lambda^2j^2/s}
\end{equation}
that was used in the calculation of \cite{MZ06} for DDBCs and  NNBCs at ${d=3}$. The scaling functions $\Upsilon_d^{\text{DD}}(x_\lambda,x_\xi)$ and $\Upsilon_d^{\text{NN}}(x_\lambda,x_\xi)$ can be computed for $2<d\ne3$ along similar lines by expanding the analog of the exponential in the first line of Eq.~\eqref{eq:Upsperd}, integrating termwise,  and using the Jacobi identity~\eqref{eq:Jacobid}. From the first term on the right-hand side of Eq.~\eqref{eq:Jacobid} the bulk term $\varphi_{\text{b},d}$ is recovered. The second one yields surface contributions $2\varphi^{\text{D}}_{\text{s},d}(T,\mu)=-2\varphi^{\text{N}}_{\text{s},d}(T,\mu)$ that are in accordance with Eqs.~\eqref{eq:varphisD} and \eqref{eq:varphisN}, respectively. The last term on the right-hand side of Eq.~\eqref{eq:Jacobid} yields the expressions for the scaling functions $\Upsilon_d^{\text{DD}}(x_\lambda,x_\xi)$ and $\Upsilon_d^{\text{NN}}(x_\lambda,x_\xi)$  given in Eq.~\eqref{eq:UpsDDNN}.

We refrain from rewriting the double series expansion for $\Upsilon_d^{\text{DN}}$ one obtains from the analog of Eq.~\eqref{eq:Upsperd} with the aid of the Poisson summation formula~\eqref{eq:Poissum} because we can get both $\Upsilon_d^{\text{DN}}$ as well as $\Theta_d^{\text{DN}}$ upon setting $(\scc_1,\scc_2)=(\infty,0)$ in the results for RBCs given in Eqs.~\eqref{eq:UpsRresd} and \eqref{eq:ThetaR}. Integrating by parts  the corresponding expression implied by Eq.~\eqref{eq:ThetaR} yields Eqs.~\eqref{eq:ThetadDNa} and \eqref{eq:ThetadDNb}.

\section{Consistency of Eqs.~\eqref{eq:fsRint} and \eqref{eq:fs2}\label{app:fsclassequiv}}

In this Appendix, we show the consistency of  Eqs.~\eqref{eq:fsRint}, \eqref{eq:Jd}, and \eqref{eq:fs2} by rederiving the first of these equations from the last one. Equation~\eqref{eq:fsRint} involves an integral $\int\rmd{k}\ldots$ over an even function of $k$, which converges for $d<2$ and is defined by analytic continuation for $2\le d<4$. Upon changing to the integration variable $E=k^2$, we can rewrite it as a contour integral along the contour $\mathcal{C}_1$ depicted in Fig.~\ref{fig:fig3DR16b} so that Eq.~\eqref{eq:fsRint} becomes
\begin{eqnarray}\label{eq:fsRmod}
\frac{f^{\text{R}}_{\text{s},d}(\xi,c)}{\Gamma(1/2-d/2)}&=&\strut  \int_{\mathcal{C}_1}\frac{\rmd{E}}{8\pi E}\left(\frac{E+\xi^{-2}}{4\pi}\right)^{(d-1)/2}\frac{2c\sqrt{E}}{E+c^2}\nonumber\\ &&\strut +
\frac{1}{4}(4\pi\xi^2)^{(1-d)/2}.
\end{eqnarray}
Let us  add and subtract $\rmi(E-c^2)$ to the numerator $2c\sqrt{E}$ of the fraction. The subtracted term yields a contribution proportional to the residue at $E=0$ of the associated integrand. It cancels the last term in Eq.~\eqref{eq:fsRmod}. The remaining integral can be rewritten to obtain
\begin{eqnarray}\label{eq:fsRmod2}
\frac{f^{\text{R}}_{\text{s},d}(\xi,c)}{\Gamma(1/2-d/2)}&=&\strut \int_{-1/\xi^2}^{-\infty}\frac{\rmd{E}}{ 4\pi E}\bigg\{\left(\frac{-E-\xi^{-2}}{4\pi}\right)^{(d-1)/2}\nonumber\\ &&\times \frac{E^{1/2}-\rmi c}{E^{1/2}+\rmi c}\sin[\pi (d-1)/2]\bigg\}\nonumber\\&=&c^{d-1}\frac{K_{d-1}}{\Gamma[(1-d)/2]}\,J_d(\xi c).
\end{eqnarray}
The result given in the last line  is equivalent to  Eqs.~\eqref{eq:fsRint} and \eqref{eq:Jd}. To get it we transformed to the integration variable $p=\sqrt{-E-1/\xi^2}$.

\section{Scaling functions of the residual free energy and residual grand potential for Robin boundary conditions\label{app:UpsR}}

Here, we complete our calculations of the scaling functions $\Theta_{d}^{\text{R}}(x_\xi,\scc_1,\scc_2)$ and $\Upsilon_{d}^{\text{R}}(x_\lambda,x_\xi,\scc_1,\scc_2)$, establishing the results given in Eqs.~\eqref{eq:ThetaR} and \eqref{eq:UpsRresd}. We first describe an alternative, somewhat easier way of computing these functions. 

The function $\Theta_{d}^{\text{R}}$  was computed for ${x_\xi=0}$ in \cite{SD08,EGJK08}, a value for which it reduces to the scale-dependent Casimir amplitude of Eq.~\eqref{eq:DeltaRdef}. The calculation used in these references can be  extended in a straightforward fashion to the noncritical case $\xi_\infty>0$  to derive the result given in Eq.~\eqref{eq:ThetaR}.  A convenient alternative way is to consider an $O(2n)$ massive free field theory in the infinite space $\mathbb{R}^d$, impose the boundary conditions~\eqref{eq:BCR} via $\delta$~functions, represent these  $\delta$~functions as integrals over  auxiliary fields $\chi_1(\bm{y})$ and $\chi_2(\bm{y})$ with support on the planes ${z=0}$ and $D$, respectively, and integrate out $\bm{\phi}$ (see, e,g., \cite{EGJK08,LK92,BDS10}). This gives a Gaussian free energy from which we subtract its value for $D=\infty$, obtaining
\begin{equation}
n\int\frac{\rmd^{d-1}p}{(2\pi)^{d-1}}\ln \det\left\{ \frac{\mathcal{M}_D[\kappa_p(\xi),c_1,c_2])}{\mathcal{M}_\infty[\kappa_p(\xi),c_1,c_2]}\right\},
\end{equation}
where 
\begin{equation}
\mathcal{M}_D(\kappa,c_1,c_2)=\begin{pmatrix}
\frac{c_1^2-\kappa^2}{2\kappa}&\frac{(c_1-\kappa)(c_2-\kappa)}{2\kappa}\,\rme^{-\kappa D}\\[\medskipamount]
\frac{(c_1-\kappa)(c_2-\kappa)}{2\kappa}\,\rme^{-\kappa D}&\frac{c_2^2-\kappa^2}{2\kappa}
\end{pmatrix}
\end{equation}
Going over to scaled variables then gives the result reported in Eq.~\eqref{eq:ThetaR}.

An analogous procedure can be used to compute the scaling function $\Upsilon_{d}^{\text{R}}(x_\lambda,x_\xi,\scc_1,\scc_2)$. Let  $S_0^\infty[\bm{\psi}^*,\bm{\psi}]$ be the coherent-state action of a free massive $n$-component quantum theory defined by Eq.~\eqref{eq:S0}  with $\mathfrak{V}=\mathbb{R}^d$ and consider the restricted partition function 
\begin{equation}
\Xi^{\text{R},\infty}_0(T,\mu,c_1,c_2)\equiv\int\mathcal{D}_{\text{R}}[\bm{\psi}^*,\bm{\psi}]\,\rme^{-S_0^\infty[\psi^*,\psi]/\hbar}
\end{equation}
with 
\begin{eqnarray}
\lefteqn{\mathcal{D}_{\text{R}}[\bm{\psi}^*,\bm{\psi}]}&&\nonumber\\ &=&\mathcal{D}[\bm{\psi}^*,\bm{\psi}]
\prod_{\bm{y},\tau,\alpha}\prod_{j=1}^2\big\{\delta[(\partial_z+(-1)^jc_j)\psi_\alpha(\bm{y},D\delta_{j,2},\tau)]\nonumber\\ &&\times\delta[(\partial_z+(-1)^jc_j)\psi^*_\alpha(\bm{y},D\delta_{j,2},\tau)]\big\}.
\end{eqnarray}
Here the $\delta$~functions ensure that the fields $\bm{\psi}(\bm{y},z,\tau)$ and $\bm{\psi}^*(\bm{y},z,\tau)$ satisfy the  RBCs~\eqref{eq:BCR}. Representing these $\delta$~functions by means of two pairs of $n$-component auxiliary fields $\bm{\chi}_j(\bm{y},\tau)$ and $\bm{\chi}^*_j(\bm{y},\tau)$  located on the respective planes $z=D\delta_{j,2}$, we arrive at
\begin{eqnarray}
\lefteqn{\Xi^{\text{R},\infty}_0(T,\mu,c_1,c_2)=\int_{\bm{\chi}^*,\bm{\chi},\bm{\psi}^*\bm{\psi}}\rme^{-S_0^\infty[\psi^*,\psi]/\hbar}}&& \nonumber\\&&\times 
\prod_{j=1}^2\exp\bigg[\rmi\int_0^{\beta\hbar}\rmd{\tau}\int\rmd^{d-1}y\big\{\bm{\chi}^*_j\cdot\big(\partial_z+(-1)^jc_j\big)\bm{\psi}\big|_j\nonumber\\&&\strut +\bm{\chi}_j\cdot\big(\partial_z+(-1)^jc_j\big)\bm{\psi}^*\big|_j\big\}\bigg],
\end{eqnarray}
where $\int_{\bm{\chi}^*,\bm{\chi},\bm{\psi}^*\bm{\psi}}$ is a short-hand for the functional integrals $\int\mathcal{D}[\bm{\chi}^*,\bm{\chi}]\int\mathcal{D}[\bm{\psi}^*,\bm{\psi}]$ and $|_j$ means that $z$ has been set to $D\delta_{j,2}$.

The action is quadratic in the fields $\bm{\psi}^*\ldots,\bm{\chi}$. We can first perform the functional integration $\int_{\bm{\psi}^*,\bm{\psi}}$ and subsequently $\int_{\bm{\chi}^*,\bm{\chi}}$. The integrand of the latter integral is a Gaussian involving the matrix kernel
\begin{align}
&\begin{pmatrix}
(\partial_n-c_1){}_1|g^{(\text{b})}|_1(\overleftarrow{\partial_n}-c_1)&(\partial_n-c_1){}{}_1|g^{(\text{b})}|_2(\overleftarrow{\partial'_n}-c_2)\\
(\partial_n-c_2){}_{2}|g^{(\text{b})}|_1(\overleftarrow{\partial'_n}-c_1)&(\partial_n-c_2){}_{2}|g^{(\text{b})}|_2(\overleftarrow{\partial'_n}-c_2)
\end{pmatrix}\nonumber\\ 
&=\frac{2m}{\hbar}\bm{\mathcal{M}}_D[\kappa_{p,\rho}(\xi,\lamth),c_1,c_2],
\end{align}
where $\partial_n$ means the inner normal, i.e., $\partial_n|_1=\partial_z|_{z=0}$ and $\partial_n|_n=-\partial_z|_{z=D}$. The function $g^{(\text{b})}$ denotes the (free bulk) propagator 
\begin{eqnarray}
\lefteqn{\langle\psi_\alpha(\bm{y},z,\tau)\psi^*_\beta(\bm{0},0,0)\rangle_0^{(\text{b})}}&&\nonumber \\ &=&\delta_{\alpha\beta}\pint\rme^{\rmi\bm{p}\cdot\bm{y}}\frac{1}{\beta\hbar}\sum_{\rho=-\infty}^\infty g^{(\text{b})}(\bm{p},\rho,z)\,\rme^{-\rmi\omega_\rho\tau},
\end{eqnarray}
associated with the action $S_0^\infty$ and is given by
\begin{equation}
g^{(\text{b})}(\bm{p},\rho,z)=\frac{2m}{\hbar}\,\frac{1}{2\kappa_{p,\rho}(\xi,\lamth)}\,\rme^{-\kappa_{p,\rho}(\xi,\lamth)|z|}.
\end{equation}
Performing the Gaussian integral $\int_{\bm{\chi}^*,\bm{\chi}}$ yields a determinant for the ratio of the partition functions $\Xi^{\text{R},\infty}_0(T,\mu,c_1,c_2)=\int\mathcal{D}_{\text{R}}[\bm{\psi}^*,\bm{\psi}]\,\rme^{-S_0^\infty/\hbar}$ and $\Xi^\infty_0(T,\mu)=\int\mathcal{D}[\bm{\psi}^*,\bm{\psi}]\,\rme^{-S_0^\infty/\hbar}$. The value  of its logarithm at $D=\infty$ is easily subtracted. One thus arrives at
\begin{align}\label{eq:varphiresalt}
&\varphi^{\text{R}}_{\text{res},d}(T,\mu,c_1,c_2)\nonumber \\ &=D^{1-d}\sum_\rho K_d\int_0^\infty\rmd{p}\,p^{d-2}\,g_{\scc_1,\scc_2}[\kappa_{p,\rho}(1/x_\xi,1/x_\lambda)].
\end{align}
The $\rho=0$ contribution yields the limiting classical behavior. The contributions for $\rho=\pm |\rho|\ne0$ yield the sum of quantum corrections resulting from the second term in the curly brackets of Eq.~\eqref{eq:UpsRresd}.

What remains to show is the equivalence of  Eq.~\eqref{eq:varphiRres} with Eq.~\eqref{eq:UpsRresd}. We use the fact that $T_{\text{res}}(z;1,\scc_1,\scc_2)$ can be written as
\begin{equation}
T_{\text{res}}(\zeta;1,\scc_1,\scc_2)=\partial_\zeta\ln\frac{R_{\scc_1,\scc_2}(\sqrt{\zeta})}{R^{(0)}_{\scc_1,\scc_2}(\sqrt{\zeta})}
\end{equation}
with
\begin{equation}
R^{(0)}_{\scc_1,\scc_2}(\sqrt{\zeta})=\frac{(\sqrt{\zeta}+\rmi\scc_1)(\sqrt{\zeta}+\rmi\scc_2)}{2\rmi\sqrt{\zeta}}\,\rme^{-\rmi\sqrt{\zeta}},
\end{equation}
so that
\begin{equation}
\frac{R_{\scc_1,\scc_2}(\sqrt{\zeta})}{R^{(0)}_{\scc_1,\scc_2}(\sqrt{\zeta})}=1-\frac{(\sqrt{\zeta}-\rmi\scc_1)(\sqrt{\zeta}-\rmi\scc_2)}{(\sqrt{\zeta}+\rmi\scc_1)(\sqrt{\zeta}+\rmi\scc_2)}\,\rme^{\rmi2\sqrt{\zeta}}.
\end{equation}
Using these results to express the function $T_{\text{res}}$ in Eq.~\eqref{eq:varphiRres}, one can integrate by parts and transform to the variable $p=\sqrt{-u}$. One thus arrives at Eq.~\eqref{eq:UpsRresd}.

%


\begin{thebibliography}{59}
\bibitem [{fn1({\natexlab{b}})}]{fn1}%
  \BibitemOpen
  \href@noop {} {}\bibinfo {note}  {Note that $\varphi
  _{\protect \mathrm{b},d}$ and $\varphi^{\protect \text{BC}}_{\protect
  \text{s},d}$ are defined through appropriate limits $D\to \infty $ at fixed
  $\lambda _{\protect \text{th}}$. If $\lambda _{\protect \text {th}}\gg D$, a
  limit recently considered in Ref.~\cite {JNS16}, then the decomposition
  looses its significance.}\BibitemShut {Stop}%
\bibitem [{fn2()}]{fn2}%
  \BibitemOpen
  \href@noop {} {}\bibinfo {note} {In the case of interacting Bose gases, there
  are additional length scales associated with the interaction; see, e.g.,
  Ref.~\cite {WRFS86}. Likewise, additional lengths related to surface
  properties, such as $1/c_1$ and $1/c_2$, come into play if one considers
  films subject to Robin boundary conditions.}\BibitemShut {Stop}%
\bibitem [{fn3()}]{fn3}%
  \BibitemOpen
  \href@noop {} {}\bibinfo {note} {This function differs from the function
  $\protect \mathcal {R}_{\protect \mathsf {c}_1,\protect \mathsf
  {c}_1}(\protect \mathsf {k})$ used in Refs.~\cite {SD08}, \cite {DS11}, and
  \cite {DrFSchmidt}. Our different choice here serves to avoid a zero at
  $\protect \mathsf {k}=0$ whenever $\protect \mathsf {c}_1$ and $\protect
  \mathsf {c}_2$ do not both vanish.}\BibitemShut {Stop}%
\bibitem [{fn1({\natexlab{c}})}]{fn10}%
  \BibitemOpen
  \href@noop {} {} \bibinfo {note} {It has been known for
  long that appropriate classical models describing the critical behavior of
  quantum models near critical temperatures $\tc>0$ result upon replacement of
  $\bm{\psi}$ by its zero-matsubara frequency component; see, e.g.,
  \cite{Her76}. The fact that the classical $2n$-component $\phi^4$ model
  results in this manner was used, for example, in the bulk case to compute the
  interaction-induced shift of the bulk critical point by means of a $1/n$
  expansion applied to the $\phi^4$ theory; see, e.g., \cite{BBZ-J00},
  \cite{Kas04}, and their references.}\BibitemShut {Stop}%
\bibitem [{fn4()}]{fn4}%
  \BibitemOpen
  \href@noop {} {}\bibinfo {note} {The two expansion terms given for ${d=3}$
  can be obtained by expanding the two expansion terms given for ${d\ne }3$ to
  order $x \ln x$. The pole $\propto x^2/(d-3)$ that the Laurent series of the
  $x^{(d-1)/2}$ term yields is of order $x^2$.}\BibitemShut {Stop}%
\bibitem [{fn5()}]{fn5}%
  \BibitemOpen
  \href@noop {} {}\bibinfo {note} {We use dimensional regularization for the
  classcial theory. As is expounded in Ref.~\cite{DGHHRS14}, the value $\infty$
  may be viewed as a fixed-point value of $g$. Setting $g=\infty$ eliminates
  the Wegner-type corrections to scaling associated with deviations of $g$ from
  its fixed-point value.}\BibitemShut {Stop}%
\bibitem [{fn6()}]{fn6}%
  \BibitemOpen
  \href@noop {} {}\bibinfo {note} {Note that the second line of Eq.~(4.72) of
  Ref.~\cite{DDG06} contains a misprint: the right-most
  $\rme^{\sqrt{\check{r}_\infty/}2}$ should be replaced by
  $\rme^{\sqrt{\check{r}_\infty}}$.}\BibitemShut {Stop}%
\bibitem [{fn7()}]{fn7}%
  \BibitemOpen
  \href@noop {} {}\bibinfo {note} {The result given here follows from the
  result given in Eq.~(A3) of \cite{DGHHRS14} for the integral
  $I_d({r=1/\xi^2})$ by multiplying it with $-A_{d-1}$ and the use of a
  standard transformation formula for the hypergeometric function.}\BibitemShut
  {Stop}%
\bibitem [{fn8()}]{fn8}%
  \BibitemOpen
  \href@noop {} {}\bibinfo {note} {As can be seen from Eq.~\eqref{eq:u0d}, the
  coupling constant $\ub$ of the (${d=3}$)-dimensional Bose theory is linear in
  the $s$-wave scattering length $a_s$. its (${d=2}$)-dimensional counterpart
  vanishes (see Ref.~\cite{BDZ08} and its references). In the case of a
  strictly two-dimensional interacting Bose gas, one would have to choose an
  energy-dependent interaction constant \cite{BDZ08}.}\BibitemShut {Stop}%
\end{thebibliography}
\end{document}